# To What Extent Do Disadvantaged Neighborhoods Mediate Social Assistance Dependency? Evidence from Sweden

Cheng Lin[1]*, Adel Daoud[1,2], Maria Brandén[1,3]

1. Institute for Analytical Sociology, Department of Management and Engineering, Linköping University, Norrköping, Sweden

2. The Division of Data Science and Artificial Intelligence for the Social Sciences, Department of Computer Science and Engineering, Chalmers University of Technology, Gothenburg, Sweden

3. Department of Sociology, Stockholm University, Stockholm, Sweden

* Corresponding author: cheng.lin@liu.se

**Abstract**

Occasional social assistance prevents individuals from a range of social ills, particularly unemployment and poverty. It remains unclear, however, how and to what extent continued reliance on social assistance leads to individuals becoming trapped in social assistance dependency. In this paper, we build on the theory of cumulative disadvantage and examine whether the accumulated use of social assistance over the life course is associated with an increased risk of future social assistance recipiency. We also analyze the extent to which living in disadvantaged neighborhoods constitutes an important mechanism in the explanation of this association. 17-year Swedish register data allow us to apply causal mediation analysis, and thereby quantify the extent to which the likelihood of ending up in social assistance dependency is affected by residing in disadvantaged neighborhoods. Our findings show the accumulation of social assistance over the studied period is associated with a more than the four-fold increase on a risk ratio scale for future social assistance recipiency, compared to never having received social assistance during the period examined. Our results suggest that the indirect effect of disadvantaged neighborhoods is weak to moderate. Therefore, social assistance dependency may be a multilevel process.

**Acknowledgments and Credits** This research was supported by grants from The Swedish Research Council (VR) grant no. 2019-00245, 2013-07681, 2020-01285 and Riksbankens Jubileumsfond (RJ) grant no. M18-0214:1. We thank participants in INAS 14th Annual Conference in Florence, EpiDem Network Junior Conference in Sweden, ECSR-CCA-NASP Spring School 2022 in Turin, and Lyckliga gatan workshop at Stockholm University for helpful feedback and discussions.

**Introduction**

The redistribution of wealth from the richest part of a population to the most disadvantaged constitutes an important means of improving equality in opportunities (Birnbaum 2010; Rawls 1971). Even though research has identified benefits from social assistance (REF), there are concerns about its possible side effects, with prolonged dependence on assistance (also known as, *recidivism*) being one such concern. It has been suggested that long-term or repeated reliance on social assistance may lead to a loss of human capital and have negative effects on individuals' socioeconomic status (Mood 2013); we refer to this reliance as *social assistance dependency*: a person's reliance on social assistance. The greater this reliance, the higher the degree of dependency. Individual incentives and generous social policies are often considered drivers of welfare dependency (Dahl, Kostøl, and Mogstad 2014). In contrast, social assistance dependency may be a process of cumulative disadvantage resulting from inadequate policy design. In Sweden, a portion of the wealth that is redistributed in society flows to the most disadvantaged individuals via social assistance, which is a direct, means-tested form of financial assistance for those living below a minimum poverty threshold. However, levels of Swedish social assistance are not sufficient to fundamentally reduce poverty, which provides an opportunity to explore social assistance dependency as a process of cumulative disadvantage, rather than only on the basis of a traditional behavioral perspective (Hussénius 2021; Korpi and Palme 1998). A lack of knowledge in this area is likely to lead to a substantial misallocation of welfare resources, since approximately SEK 11.9 billion was paid in social assistance in 2020.[1] In the US, federal means-tested programs accounted for USD 148 billion in cash assistance provision.[2]

---

[1] https://www.socialstyrelsen.se/globalassets/sharepoint-dokument/artikelkatalog/statistik/2021-6-7469.pdf
[2] https://poverty.ucdavis.edu/sites/main/files/file-attachments/43935-means-tested-infographic.pdf

Although much research exists on the determinants of individual poverty and on how social assistance alleviates poor living conditions, less is known about how the continued use of such assistance may create social assistance dependency—perhaps aggravating poverty later in life. Some studies have speculated that such dependency exists. For example, Mood (2013) and Bäckman and Bergmark (2011) speculated on the existence of dependency, although the estimated effect size of social assistance was relatively small and dependent on selection into social assistance. Qualitative research shows that social assistance benefits in Sweden help disadvantaged individuals in the short term, but are less helpful in the long run (Marttila et al. 2010). Similar results have been found in Canada (Cooke 2009), the EU (Königs 2018), and the US (Meyer 1990). However, even though these studies suggest that some individuals experience dependency during a certain period of the life course, there is a lack of knowledge about its cumulative effect over longer periods of time. A *cumulative effect of social assistance recipiency* would involve the accumulated amount of past social assistance leading to a higher probability of an individual receiving even more social assistance in the future.

The determinants of social assistance dependency are both behavioral and contextual, but it remains unclear which determinants are most significant over the course of a person's life. Social assistance dependency has been discussed from a behavioral perspective, with researchers suggesting that the primary drivers of such dependency are incentives, generous social policies, and individual-level behaviors (Dahl et al. 2014). From a structural perspective, individuals' life-course outcomes and economic behaviors are embedded in the context, which provides resources and opportunities (Brady 2019). The concentration of poverty in disadvantaged neighborhoods could thus constitute a determinant of poverty (Brady 2019; Harding 2010).

We argue, however, that social assistance dependency should be understood as a process of cumulative disadvantage, or more specifically a chain-of-risk, whereby individuals on social assistance become increasingly exposed to other adverse factors—both behavioral and contextual—which in turn increase their likelihood of again becoming reliant on social assistance. In particular, we argue that contextual-level characteristics, specifically disadvantaged neighborhoods, are likely to be important predictors of reliance on social assistance (Gustafsson, Katz, and Osterberg 2019). For example, unequal housing markets and poverty are likely to increase social assistance recipients' risk of having to remain in, or move to, disadvantaged neighborhoods, which would in turn give them less access to favorable network structures and employment opportunities. The lack of such attributes in disadvantaged neighborhoods is likely to increase the probability of social assistance recipiency and is a potentially crucial mediating factor for understanding the process of social assistance dependency. We expect that expanding our theorizing of social assistance dependency to include structural components of this kind will improve our understanding of this phenomenon. If neighborhoods play a major role in mediating social assistance dependency, this would indicate that the structured concentration of poverty plays a substantial role in restricting individuals' actions and choices. If neighborhoods play only a small role, or have no statistically and substantively discernible effects in the mediation of social assistance dependency, then this small (or null) effect of concentrated poverty would instead indicate that dependency is likely more to be affected by individual-level factors, i.e. motivation, opportunities or other demographic factors.

To evaluate these alternatives, we use Swedish register data to follow a cohort born in 1981 from age 20 to age 36. We measure annual social assistance recipiency and individualized neighborhood disadvantage defined on the basis of the characteristics of the individuals living around each ego. We then estimate the direct effect of cumulative prior social assistance recipiency and the indirect effect of cumulative exposure to disadvantaged neighborhoods on social assistance recipiency in mid-adulthood. Estimating effects based on a time-varying setting of this kind poses difficult methodological challenges in terms of both unobserved confounders and biases due to the presence of treatment-induced mediator-outcome confounders. We partially overcome these challenges by relying on *interventional effects* that mean interventions on treatment and/or mediator in the mediation setting (Nguyen, Schmid, and Stuart 2021; VanderWeele, Vansteelandt, and Robins 2014), and a sensitivity analysis method. By combining Sweden's rich population registers with this innovative methodological strategy, we are able to examine the relationship between neighborhoods and social assistance dependency over a considerable period of a person's life history—something that has previously been impossible to observe.

**Literature Review and Conceptual Framework**

**1. Social Assistance and Cumulative Disadvantage**

The use and dynamics of social assistance are contextually dependent (Gustafsson et al. 2019). Social assistance recipients tend to receive assistance for spells of longer duration in the Netherlands and Luxembourg, as compared to Norway and Sweden where social assistance recipients are more likely to have multiple, but briefer, spells of recipiency (Königs 2018). Research addressing social assistance dependency has tended to focus on examining spells of social assistance recipiency and to explore the duration or state dependence of such spells. Time-to-event analyses and analyses of within-spell durations (Bäckman and Bergmark 2011; Mood 2013) have, for instance, shown an increased risk of remaining on social welfare by the duration of the period of recipiency. In addition, most previous research has restricted its focus to examining only the first observed spell of recipiency. This is unfortunate given that the relationship between earlier and later spells of social assistance recipiency might itself also constitute a variant of welfare dependency (Blank 1989; Mood 2013) that we to date know little about. Given the high incidence of repeated spells of social assistance (Königs 2018), a failure to take multiple spells into account will at best provide only an incomplete picture of social assistance dependency, as has been argued by Tseng, Vu, and Wilkins (2008). From the US policy debate, there has been a concern that financial support directed at the poor could lead to welfare dependency[3]. However, research has produced contrasting results (Baird, McKenzie, and Özler 2018). Banerjee et al. (2017) did not find that cash transferences reduced work incentives in experimental settings in six developing countries. Further, Aizer, Hoynes, and Lleras-Muney (2022) found that the provision of means-tested assistance as a social safety net could benefit children by reducing child poverty. A possible explanation is that non-generous social assistance benefits may not alone be sufficient to support disadvantaged individuals by alleviating poverty, which is produced by the process of cumulative disadvantage. In contrast to previous research, Markussen and Røed (2016) found that a generous income support program with activation requirements, which provided a yearly salary twice as high as that expected on social assistance, had positive effects on employment in Norway.

The link between past and future social assistance recipiency can be understood through the framework of cumulative disadvantage. Cumulative disadvantage describes the process by which the previous disadvantage increases the risk of subsequent disadvantage. Berg (2017) has argued that cumulative disadvantage might be modelled in terms of (1) cumulation or (2) a chain of risk (this is also discussed by Ben-Shlomo, Mishra, and Kuh (2014) from a life-course perspective). The cumulative model highlights the effects of repeated exposures on the outcome. For instance, Wodtke, Harding, and Elwert (2011) found that cumulative exposure to the most disadvantaged neighborhoods could decrease the likelihood of high school graduation among black students from 96% to 76%. The chain-of-risk model instead emphasizes the importance of mediation. For example, Wodtke, Ramaj, and Schachner (2022) show that lead contamination may be a key mechanism in understanding the relationship between disadvantaged neighborhoods and low vocabulary ability.

---

[3] https://epod.cid.harvard.edu/article/dispelling-myth-welfare-dependency

In this article, we use the modified *Chain of Risk Additive Model* (see the graphical model in Section 3) as described by Ben-Shlomo, Mishra, and Kuh (2014), to understand social assistance dependency as a process of cumulative disadvantage. This model captures both cumulative characteristics and mediating properties. In Figure 1, we present a 3-period modified *chain of risk additive model* to clarify how social assistance recipiency at time 1 and 2 cumulatively impacts social assistance recipiency at time 3 via the mediating process of exposure to disadvantaged neighborhoods at time 1 and 2. The arrows indicate relationships between different variables. In addition, as noted by Lynn and Espy (2021), initial endowment and system-imposed constraints may constitute key mechanisms for understanding cumulative disadvantage. Disadvantaged neighborhoods could harm individuals during early adulthood (Gustafsson et al. 2019), which could lead to the worse initial endowment. If so, disadvantaged neighborhoods might constitute a system-imposed constraint, since they may produce unequal exposure to worse environments, e.g., air pollution, which may in turn harm long-term human capital investment (Wodtke et al. 2020), and lead to structural disadvantages for the poor in disadvantaged neighborhoods.

*[Figure 1 about here]*

An additional process through which cumulative disadvantage operates involves scarring effects (Birkelund, Heggebø, and Rogstad 2017; DiPrete and Eirich 2006), a term that refers to the way that experiences of disadvantage, however long ago they occurred, increase the probability of future disadvantage (Ilmakunnas and Moisio 2019), for example through their negative impact on human capital accumulation (DiPrete and Eirich 2006; Pissarides 1992). The prevalence of scarring effects has primarily been studied in relation to unemployment, with research showing causal scarring effects from early-career unemployment on future unemployment patterns (Schmillen and Umkehrer 2017). Ilmakunnas and Moisio (2019), suggest that this is the case also for social assistance. Their findings indicate that receiving social assistance in young adulthood increases the risk for later social assistance recipiency due to its scarring effect. This interpretation is supported by findings showing that unemployment is a crucial predictor of later social assistance recipiency (Mood 2013).

Building on previous literature, we expect to find the dependence between past and future social assistance recipiency, whereby the cumulative number of years on social assistance increases the risk for future social assistance. Thus, we formulate our first hypothesis as:

*H1: The cumulative number of years with social assistance recipiency between the ages 20 and 33 will increase the risk of social assistance recipiency at age 34-36.*

## 2. Neighborhood Disadvantage as a Mechanism Explaining Social Assistance Dependency

A common characteristic for almost all studies on the determinants of social assistance dependency is their focus on the individual-, parental-, or household-level characteristics. Previous research has shown that social assistance recipiency and dependency are highly dependent on socioeconomic backgrounds (Mood 2013), parental characteristics and other individual-level attributes, such as low education – both one's own (Hümbelin and Fritschi 2018; Ilmakunnas and Moisio 2019) and one's parents' (Ilmakunnas 2018), socioeconomic status, possible criminal activity (Stenberg 2000), immigrant background (Carpentier, Neels, and Van den Bosch 2017), and financial vulnerability and unemployment (Contini and Negri 2007; Stranz and Wiklund 2012). There are some exceptions however. Åslund and Fredriksson (2009) examined the importance of residential context and found that living in communities with high levels of welfare dependency increased individuals' own risk of long-term welfare dependency. They found that a 10% increase in exposure to peer welfare dependency implied a 0.9% increase in individuals' probability of welfare recipiency. Markussen and Røed (2016) have also examined peer effects on social insurance use in Norway. With these exceptions, research has neglected to acknowledge the importance of meso-level characteristics, such as those related to neighborhoods, in studies on social assistance recipiency. We would argue, however, that the neighborhood constitutes a mechanism that is likely to operate in the intersection between initial endowment and system-imposed constraints, which constitute two fundamental mechanisms of cumulative disadvantage (Lynn and Espy 2021).

Importantly, even if some studies suggest that exposure to a disadvantaged neighborhood increases the risk for social welfare recipiency (Mood 2010), no study has linked this possibility to the overexposure of vulnerable individuals to such neighborhoods and examined how this in turn impacts social assistance dependency. We argue that there are two distinct processes that contribute jointly to this pattern: (1) the selection of individuals with prolonged periods of social assistance recipiency to neighborhoods with adverse characteristics, and (2) the effects of such adverse characteristics on social assistance dependency. Both of these processes are consistent with theories of cumulative disadvantage, and we discuss them in detail below.

**2.1 How Cumulative Social Assistance Recipiency Impacts Neighborhood Selection**

When studying neighborhood effects, selection into neighborhoods is often viewed as a nuisance factor that needs to be adjusted for. However, neighborhood selection can also constitute an important mechanism that increases our understanding of a certain phenomenon (Arcaya et al. 2014; Sampson 2011; Sampson and Sharkey 2008). It is reasonable to assume that prolonged or repeated social assistance recipiency is likely to increase the probability of remaining in or moving to neighborhoods that are less beneficial to live in, which is in turn likely to impact individuals' future life prospects, such as their likelihood of relying on social assistance in the future. One example of such a selection process has been described by Arcaya et al. (2014), who demonstrated that individuals' health status impacted where they live through the positive association between poor health and neighborhood poverty.

In line with the Desires-Beliefs-Opportunities framework (Hedstrom 2005), we argue that individuals' residential choices stem from what they *desire* from a neighborhood, their *beliefs* about a certain neighborhood, and the *opportunity* they have to live in a given neighborhood. Most importantly, these three dimensions, and particularly the opportunities dimension, will differ by levels of previous social assistance recipiency.

A possible explanation why low-income families or individuals tend to live in low-opportunity areas is their preference for affordability or for living close to family and work opportunities (Bergman et al. 2019). However, neighborhood selection processes are not only driven by the preferences of low-income families but also by those of high-income families.

Schachner and Sampson (2020) found that parents with high cognitive ability were almost 50% more likely to select an advantaged area. In addition, Reardon and Bischoff (2011) have noted three potential income-related residential preferences: neighbors' SES, neighbors' income-correlated characteristics, and local public goods. Furthermore, Reardon and Bischoff (2011) also found that residential segregation between the highest-income families and others constitutes the key process in the way income inequality impacts income segregation, by inducing the highest-income households to move away from others. Affluent residents tend to avoid disadvantaged neighborhoods, or leave them as the share of non-affluent residents increases (Benard and Willer 2007), which would further increase the negative sorting of social assistance dependency (see Fjellborg 2021 for research showing similar results in Stockholm, Sweden).

In terms of opportunities, Reardon and Bischoff (2011) argue that housing policy may be a central determinant of the segregation between the poor and the non-poor in the US. There are also likely to be barriers that prevent low-income households from leaving low-opportunity areas, a view that is supported by Bergman et al. (2019), who found that 53% of families who were offered a voucher to move to a high-opportunity area moved when they received "customized search assistance, landlord engagement, and short-term financial assistance", compared to only 15% of families who were offered a similar voucher but who did not receive the same help to move. This suggests that residential segregation by income is mainly driven by barriers associated with the residential searches of low-income families rather than their strong preferences (Bergman et al. 2019). Most importantly, therefore, residential decisions are affected by the opportunities an individual has to act on his or her residential preferences, and these opportunities are in turn highly dependent on the individual's resources (Bergman et al. 2019) and the structure of the housing market (Fjellborg 2021).

The structure of the housing market, including the distribution of rental and owned housing, housing queue systems and differences in housing costs across a city, will impact where individuals in different financial situations are able to live. As noted by Andersson and Turner (2014), central Stockholm and suburban areas are experiencing gentrification and residualization respectively, which means that the most disadvantaged individuals will become stuck in suburban areas as a result of dramatic shifts in Swedish housing policy. In the Swedish housing market, cheap housing is decreasingly often located adjacent to more expensive housing. For instance, we know that groups with low or irregular incomes are less likely to have access to newly established public housing with higher rental requirements, which exposes young adults to a cycle of resource and housing inequality, as indicated by qualitative studies by Grander (2017, 2021). Also, low-income individuals are more likely to move into disadvantaged neighborhoods in Stockholm, and the housing market in Stockholm manifests socio-spatial residential patterns that follow the distribution of incomes (Fjellborg 2021). Various strategies employed by landlords, such as raising rents, could lead to the exploitation of the disadvantaged individuals who live in disadvantaged neighborhoods, which may further exacerbate living-condition deprivation for social assistance recipients (Desmond and Wilmers 2019).

Taken together, these factors suggest that individuals who are in receipt of social assistance repeatedly or for long periods of time are more likely to face financial restrictions on where they can live, and are thus more likely to be forced to live in neighborhoods with less favorable attributes. We will now discuss how cumulatively living in disadvantaged neighborhoods might increase the risk of social assistance recipiency in the future.

## 2.2 The Effects of Residing in a Disadvantaged Neighborhood on Later Social Assistance Recipiency

Residing in a disadvantaged neighborhood is likely to affect the risk of later social recipiency for at least three reasons.

First, individuals who live in disadvantaged neighborhoods are underexposed to social interactions with high-SES neighbors (Wilson 2012). This means that they are deprived of two possibly important peer effects; that of role models and that of the transmission of information (Bolt and van Kempen 2013). In terms of role models, resourceful neighbors might improve social assistance recipients' work incentives and connections, since resourceful neighbors may also provide opportunities or information to help disadvantaged groups enter the labor market or submit job applications (Bolt and van Kempen 2013). In contrast to this argument, some qualitative studies have shown that individuals who live in low-SES neighborhoods benefit from receiving recommendations or help from each other (Pinkster 2014). These connections may be particularly useful for finding low-skilled jobs (Custers 2019). It might also be the case that regardless of neighborhood composition, social assistance recipients are more likely to make connections with other disadvantaged individuals, and less likely to receive information on opportunities for upward mobility (Browning et al. 2017; Wilson 2012), which might produce a null effect from neighborhood disadvantage.

Second, social institutions within neighborhoods can play a crucial role by providing local services for job search, school quality, and accessibility to public spaces (Galster 2012; Vandecasteele and Fasang 2020). Harjunen et al. (2021) provided empirical evidence of a decline in the number of affluent households following a large school closure in a neighborhood in Finland. Access to neighborhood amenities may be of particular importance for individuals, which may be of particular importance for individuals who are financially constrained and thus have lower access to food, medicine, and sports facilities (Camina and Wood 2009).Therefore, if individuals are living in disadvantaged neighborhoods, they may lack the supportive institutions required to maintain resourceful conditions.

Third, residing in a disadvantaged neighborhood may lead to geographical stigma (Bolt and Kempen 2013; Vandecasteele and Fasang 2020). Living in a neighborhood that has a bad reputation may harm the likelihood of getting a job because employers may evaluate applications based on where applicants live (Bunel, L'Horty, and Petit 2015). Neighborhood signaling effects may play a role in the production of geographical stigma. Findings by Carlsson, Reshid, and Rooth (2018) showed that individuals who signaled living in a disadvantaged neighborhood had 14% lower callback rates for job applications compared to individuals living in neighborhoods that were considered good. This means that welfare recipients who live in disadvantaged neighborhoods may also be less likely to exit social welfare recipiency due to geographical stigma.

In terms of neighborhood effects on social assistance, Casciano and Massey (2008) found that a 1% increase in the proportion of poor residents within US neighborhoods increased the probability of welfare recipiency among young mothers from 0.12 to 0.45, conditioned on a set of individual-level SES factors. They also found that a 1% increase in the proportion of poor residents within a neighborhood could reduce young mothers' odds of employment by 57%. In Sweden, Mood (2010) found that a higher share of social assistance recipients in the neighborhood increased the probability of a family receiving social assistance. This is in line with findings from the Netherlands (Van der Klaauw and van Ours 2003) which showed that individuals who were unemployed after leaving full-time education (single 25-year-old men without children and without previous welfare experience) had a 0.72 probability of leaving welfare if they lived in a low-unemployment area compared to a 0.54 probability if they lived in a high-unemployment area.

However, the evidence on neighborhood effects is mixed. Some previous research shows that there are no statistically discernible neighborhood effects on a range of outcomes. For example, there are no statistically appreciable effects of neighborhoods on teenage childbearing in the US or in Sweden (Galster et al. 2007; Hedman 2014). Similar results have been found for health outcomes, such as Native American infant mortality (Johnson, Oakes, and Anderton 2008) and depressive symptoms among middle-aged African Americans (Schootman et al. 2007). Some experimental research shows heterogeneity in neighborhood effects. Chetty, Hendren, and Katz (2016) found no appreciable effects of neighborhoods on educational opportunities and earnings for children who moved during the ages 13–18. In addition, publication bias could also impact the evidence of neighborhood effects, with journals being more likely to publish "significant results" rather than "insignificant results". As noted by Nieuwenhuis (2016), there is publication bias in neighborhood effect research, which may thus increase the uncertainty associated with neighborhood effect research and explain why there is less research showing a lack of statistically discernible effects (Kwan 2018).

Other studies do show statistically discernible neighborhood effects, but with varying effect sizes. Some research has found small or modest neighborhood effects. For example, Ludwig et al. (2011) found a modest long-term neighborhood effect on extreme obesity in a randomized social experiment in the US, while Galster et al. (2007), also in the US, found a non-trivial impact of neighborhood poverty on high school attainment and future earnings. In addition, a review has shown that neighborhood could only explain a small part of the differences identified in health and well-being outcomes (Jivraj et al. 2020). For neighborhood effects on children, a range of studies, both experimental and non-experimental, have shown modest neighborhood effects on children's development (Leventhal and Dupéré 2019). In contrast, some other research has shown more strong neighborhood effects. Wodtke, Harding, and Elwert (2011) found more substantial neighborhood effects on high school graduation than those reported in previous research, and in the experimental or quasi-experimental setting, appreciable neighborhood effects have been found for a range of outcomes (Chyn and Katz 2021): mortality among migrants (Deryugina and Molitor 2020), violent crime and high school dropout rates (Chyn 2018), and earnings (Collins and Wanamaker 2014).

As noted above, we use the *Chain of Risk Additive Model*, which captures both cumulative characteristics and mediating properties, to understand social assistance dependency as a process of cumulative disadvantage. In particular, we want to learn whether the neighborhood constitutes a key mechanism for understanding social assistance dependency. There is strong theoretical and empirical evidence suggesting that social assistance recipiency is likely to impact neighborhood selection through individuals' preferences and opportunities, which will in turn impact social assistance recipients' exposure to disadvantaged neighbors. Empirical research is less clear, however, about the consequences this may have for social assistance dependency. Theoretically, residing in a disadvantaged neighborhood is likely to increase the risk of future social assistance recipiency via a range of potential mechanisms, including human capital loss. Our second hypothesis therefore reads:

*H2: Cumulative exposure to disadvantaged neighborhoods mediates the relationship between prior cumulative social assistance recipiency from age 20 to 33 and late social assistance recipiency at age 34-36.*

In contrast to theory, research provides mixed empirical evidence regarding the magnitude of such an effect, and we therefore propose a sub-hypothesis which reads:

*H2a: Cumulative exposure to disadvantaged neighborhoods weakly mediates the relationship between prior cumulative social assistance recipiency from age 21 to 33 and late social assistance recipiency at age 34-36.*

## 3. Conceptual Framework and Graphical Model
### 3.1 Graphical Model

Figure 2 presents a directed acyclic graph (DAG) that encapsulates our theoretical framework. A DAG is a graphical causal model in which all directed arrows represent a causal effect from one node to another (Pearl 2014). The DAG illustrates how social assistance recipiency during young adulthood and neighborhood disadvantage impact later social assistance recipiency. In this model, $A_t$ denotes prior social assistance recipiency as treatments at time $t$. $M_t$ represents a mediator and denotes disadvantaged neighborhood exposure as mediators at time $t$. $C$ denotes all covariates during the first year of observation, also including baseline treatment and mediator values (VanderWeele and Tchetgen Tchetgen 2017). $L_t$ denotes a set of time-varying covariates at time $t$, such as homeownership. $Y$ denotes the outcome and is a binary variable indicating whether individuals are in receipt of social assistance at age 34-36. In addition, Figure 2 includes three confounders $V$, $U$, and $Z$, which are unobserved in our data. These are likely to distort the treatment-mediator (via $V$), treatment-outcome (via $U$), and mediator-outcome (via $Z$) relationships, respectively. We handle the influence of these confounders using a sensitivity analysis method—which are fully described in the *Methods* section, and whose findings are discussed under *Results.*

Figure 2 shows that past social assistance recipiency may impact future social assistance recipiency via multiple paths. First, past social assistance recipiency might impact future social assistance recipiency both cumulatively and directly. Second, past social assistance recipiency might impact future social assistance recipiency via the cumulative exposure associated with residing in a disadvantage neighborhood, which constitutes a mediating mechanism through which past social assistance recipiency operates to affect future social assistance recipiency. Third, exposure to disadvantaged neighborhoods might impact future social assistance recipiency directly.

However, Figure 2 reveals a challenge in identifying the causal effect of interest. First, there are the unobservable potential confounders $V$, $U$, and $Z$, which may bias estimates. For example, some unobservable confounders might impact the way social assistance recipiency is associated with neighborhood disadvantage. Second, there are possible treatment-induced mediator-outcome confounders (VanderWeele and Tchetgen Tchetgen 2017; VanderWeele et al. 2014). As noted by VanderWeele et al. (2014), treatment-induced mediator-outcome confounders will render the causal effect unidentified. For instance, in this study, home ownership might be a possible treatment-induced mediator-outcome confounder at time 1 (included in $L_1$ in Figure 2), being impacted by the treatment ($A_1$), but also itself impacting the relationship between disadvantaged neighborhoods as mediators ($M_2$) and social assistance recipiency at age 34-36 (the outcome, $Y$). We deal with these challenges in two ways: (1) by using interventional effects, which allow for the existence of treatment-induced mediator-outcome confounders (VanderWeele et al. 2014), and (2) by a sensitivity analysis method, which is used to analyze more exactly the impacts of unobservable confounders on our estimates.

*[Figure 2 about here]*

### 3.2 Estimands and Identification

To estimate the causal mediated effects with treatment-induced mediator-outcome confounders, we employ the interventional direct and indirect effects introduced by VanderWeele et al. (2014). There are several reasons for the use of interventional effects. First, the use of natural indirect and direct effects requires a strong assumption that that there are no mediator-outcome confounders that are affected by the exposure (VanderWeele and Tchetgen Tchetgen 2017). However, as noted above, exposure-induced mediator-outcome confounders may exist. In this situation, interventional effects constitute an alternative to the use of natural effects. Non-zero interventional effects imply that natural direct and indirect effects do exist, but we cannot gauge the precise magnitude of these natural effects (Nguyen et al. 2021). Second, interventional effects are not simply a "second-best alternative to natural indirect and direct effects (VanderWeele and Tchetgen Tchetgen 2017: 924)", but also represent a response to the policy-orientation question, since they estimate a population-level effects that is more policy-relevant effect than the individual-level effect (Nguyen et al. 2021). As noted by Demirer et al. (2021), interventional effects might in fact constitute the primary interest as a result of their practical relevance. In other words, the presence of interventional effects might suggest the importance of policy measures focused on intervening in relation to both the mediator and the exposure (Demirer et al. 2021; VanderWeele and Tchetgen Tchetgen 2017). Identifying interventional effects could thus inform more effective interventions in relation to real issues (Wodtke and Zhou 2020). For example, spatially focused measures might be a useful means of reducing social assistance dependency (Gustafsson et al. 2019).

Following the time-varying notations in Figure 2, let $M_{A=a}$ denote the value of the mediator when individuals are exposed to treatment trajectories $a$; $Y_{A=a,M=m}$ then denotes the outcome when the treatment is $a$ and mediator is $m$; and let $G_{A=a|C}$ denote the distribution of the mediator when the population is exposed to treatment trajectories $a$ conditional on covariates $C = c$. In the interventional effects, the mediator is not fixed at a certain level but is drawn from distributions (Nguyen et al. 2021). Finally, let $a$ and $a^*$ denote different treatment trajectories. With these notations, the interventional direct effect (IDE) is (Nguyen et al. 2021; VanderWeele et al. 2014; Wodtke and Zhou 2020):

$$IDE = E\left(Y_{A=a,G_{a^*|C}}\right) - E\left(Y_{A=a^*,G_{a^*|C}}\right) \qquad (1)$$

This estimand depicts the expected difference in the outcome of interest when all actors in the targeted group are exposed to treatment trajectories $a$ rather than $a^*$, and when the mediators are randomly drawn from the distribution $G$ of treatment $a^*$ conditional on covariates $C$ (Wodtke and Zhou 2020). In this situation, the interventional indirect effect (IIE) is (Nguyen et al. 2021; VanderWeele et al. 2014; Wodtke and Zhou 2020):

$$IIE = E\left(Y_{A=a,G_{a|c}}\right) - E\left(Y_{A=a,G_{a^*|c}}\right) \qquad (2)$$

This estimand depicts the expected difference in the outcome of interest when all actors are exposed to treatment trajectories $a$, and when the mediators are randomly drawn from the distribution $G$ of treatment $a$ rather than $a^*$ conditional on covariates $C$.

To identify these estimands, three assumptions are required in the longitudinal setting (VanderWeele and Tchetgen Tchetgen 2017; VanderWeele et al. 2014):

(1) $Y_{am} \perp\!\!\!\perp A_t | A_{t-1}, M_{t-1}, L_{t-1}, C$
(2) $Y_{am} \perp\!\!\!\perp M_t | A_t, M_{t-1}, L_{t-1}, C$ and
(3) $M_{at} \perp\!\!\!\perp A_t | A_{t-1}, M_{t-1}, L_{t-1}, C$

Assumption (1) requires that there are no unobserved confounders between the outcome and the treatment at time $t$ conditional on the treatment at time $t-1$, the mediator at time $t-1$, time-varying confounders at time $t-1$, and baseline confounders. Assumption (2) requires that there are no unobserved confounders between the outcome and the mediator at time $t$ conditional on the treatment at time $t$, the mediator at time $t-1$, time-varying confounders at time $t-1$, and baseline confounders. Assumption (3) requires that there are no unobserved confounders between the mediator at time $t$ and the treatment at time $t$ conditional on the treatment at time $t-1$, the mediator $t-1$, time-varying confounders at time $t-1$, and baseline confounders. The identification strategy is consistent with the DAG presented in Figure 2. Under these identification assumptions, we can link our theoretical estimands and empirical estimands to estimate effects from data (Lundberg, Johnson, and Stewart 2021). Based on these notations, estimands, and identification, the next section discusses the estimation. At the same time, in order to test how violations of these assumptions would impact results, the next section also discusses the sensitivity analysis method.

**Data and Methods**

**1. Study Population**

We use Swedish population register data containing detailed longitudinal (annual) information on individuals' social assistance uptake, other socioeconomic and demographic characteristics, and geographical characteristics. We study a full cohort of individuals who were born in 1981 and who lived in Sweden throughout the period when they were aged 20-36 (N=91,862). The focus on this cohort enables us to study individuals during both young adulthood and mid-adulthood. In Sweden, young adults are one of the groups characterized by the highest poverty risks (Lorentzen et al. 2014), which motivates our focus on this group. We start studying our sample when they are aged 20 (2001) and follow them until the year in which they turn 36 (2017).

2. Exposures, Outcome, Mediators, and Confounders

The exposure of interest in this study is *past social assistance recipiency*. It is measured annually from 2001 to 2014, when our sample was aged 20-33. Individuals who have received a positive amount of social assistance during a given year are defined as social assistance recipients in that year. While social assistance benefits in Sweden are approved and paid monthly, we only have data that are aggregated at the annual level. This means that we cannot consider status changes in social assistance recipiency within a given year. Our outcome of interest is *social assistance recipiency during mid-adulthood*, measured between 2015 and 2017 when our sample was aged 34-36. If an individual has received social assistance benefits at any point during this three-year period, he or she is defined as a social assistance recipient.

Our mediator of interest is *the level of disadvantage in an individual's neighborhood*. We use the k-nearest neighbor approach, also called individualized neighborhoods or bespoke neighborhoods, to define neighborhoods (Östh, Clark, and Malmberg 2015). This means that we define an individual's neighborhood composition based on a given number of nearest neighbors (here 50) who live around the residence of the individual in question (Malmberg and Andersson 2016). This ensures that an individual's neighborhood composition is constituted by the neighbors surrounding that individual, which reduces the risk for measurement error stemming from e.g. an individual living close to an administrative border or from other problems associated with the modifiable areal unit problem (MAUP) (Östh et al. 2015). We use coordinate data that are accurate to the level of 100 by 100 meter squares, and we create annual, individualized neighborhoods using the Geocontext (Hennerdal and Nielsen 2017). We allow an algorithm to gradually increase the circumference of a circle around each individual until the circle contains 50 adults. Thereafter, the proportion of these neighbors with certain attributes is calculated. A threshold of fifty neighbors means the inclusion of individuals with whom you are likely to share a laundry room, bike rack, garbage room, and recycling station (Amcoff et al. 2014), which makes this threshold appropriate for the present study (note however that our results are also robust to a threshold that includes the 500 nearest neighbors; see robustness checks below).

The neighborhood characteristics that we examine are: (1) the share of neighbors with only compulsory schooling, (2) the share of neighbors in the lowest income decile, (3) the share of neighbors who are social assistance recipients, (4) the share of neighbors who have been registered as unemployed at any time during the year, and (5) the share of neighbors employed in a low-skilled occupation. We use these five variables as indices to create disadvantaged neighborhood scores using principal component analysis (see Appendix A for details).

In order to adjust for possible confounders of the kind shown in the Figure 2, all analyses include a number of covariates that have previously been shown to be associated with poverty risks (Boschman et al. 2021). As noted by VanderWeele (2019), causes of the treatment, of the outcome or of both should be considered as confounders (see Appendix B for details). The time-invariant covariates included in the analysis are gender and migration background (distinguishing between foreign-born and native-born). Our time-varying covariates are measured annually from 2001 to 2014, and comprise marital status (unmarried, married/in registered partnership, divorced or widowed), a dichotomous variable indicating whether the individual lives in one of Sweden's three metropolitan counties (Stockholm, Västra Götaland or Malmö), educational level, the total number of residential moves within the country, the number of children aged 0-3, 4-6, and 7-17, unemployment status (whether the individual was registered as unemployed at any point during the year), living in non-rental housing (although we do not know whether the individual is the actual owner of the property due to data limitation), experience of working in a skilled occupation, the log of disposable income net of social assistance, parental benefits and housing benefits[4], the total amount of housing benefits received during the year, and the total amount of parental benefits received during the year. We also control for two local labor market characteristics: the number of companies in an individual's local labor market and the size of the working-age population in an individual's local labor market. Local labor markets are defined by Statistics Sweden as clusters of municipalities that have their largest commuting streams within the cluster. Because commuting patterns change over time, local labor markets are redefined annually. In 2001, there were a total of 88 local labor markets in Sweden.

*[Table 1 about here]*

## 3. Estimation Method

VanderWeele and Tchetgen Tchetgen (2017) introduced marginal structural models (MSMs) (Robins, Hernán, and Brumback 2000) with inverse probability of treatment weighting (IPTW) in order to estimate interventional direct and indirect effects. We base our analytical design on their modeling strategy. Causal mediation analysis includes a model of the outcome process and a model of the mediation process, which capture the relationships between treatment, mediator, and outcome. In our setting, outcome and treatment are binary, and the mediator is continuous. As noted by VanderWeele and Vansteelandt (2010), the advantage of the odds ratio scale for mediation analysis is that it enables researchers to use simple regression to estimate mediating effects when an outcome is both binary and rare. However, in our setting, the outcome is not rare across all strata of the treatment and mediator, which thus involves a violation of the "rare outcome assumption" (VanderWeele, Valeri, and Ananth 2019). We therefore use Poisson regression with robust standard errors to estimate the risk ratio (Chen et al. 2018; Zou 2004). The outcome model is (VanderWeele and Tchetgen Tchetgen 2017):

$$\log\{E(Y_{am})\} = \theta_0 + \theta_1 cum(a) + \theta_2 cum(m) \tag{3}$$

where $\theta_1$ is the coefficient for the cumulative total effect of prior social assistance recipiency $cum(a)$ on later social assistance recipiency over time, and $\theta_2$ is the coefficient for the cumulative total effect of neighborhood disadvantage $cum(m)$ over time. We should assume that the mediators are "multivariate normally distributed with mean $\beta_0(t) + \beta_{1t} avg(\overline{a_t})$ (VanderWeele and Tchetgen Tchetgen 2017:930) to estimate the mediation model:

$$E(M_{at}) = \beta_0(t) + \beta_{1t} avg(\overline{a_t}) \tag{4}$$

---

[4] In Swedish register data, disposable income can be negative. In such cases, we have set the log of negative disposable income to 0.

Therefore, the interventional direct effect is given by VanderWeele and Tchetgen Tchetgen (2017):

$$\log\{E(Y_{aG_{a^*}})/E(Y_{a^*G_{a^*}})\} = \theta_1\{cum(a) - cum(a^*)\} \tag{5}$$

The interventional indirect effect is also given by VanderWeele and Tchetgen Tchetgen (2017):

$$\log\{E(Y_{aG_a})/E(Y_{aG_{a^*}})\} = \theta_2 \sum_{t \leq T} \beta_{1t} [avg\{a_t\} - avg\{a^*_t\}] \tag{6}$$

For calculation simplification, we should assume that $\beta_{0t} = \beta_0$ and $\beta_{1t} = \beta_1$ are constant and that for all times $t$, $a^*_t = 0$ and $a_t = 1$. We can calculate the interventional direct effect using $T\theta_1$ and the interventional indirect effect using $T\beta_1\theta_2$ (VanderWeele and Tchetgen Tchetgen 2017).

IPTW constitutes a robust method for estimating causal functions while adjusting for time-varying confounders (VanderWeele and Tchetgen Tchetgen 2017) . IPTW is a weighting method that computes weights for individuals. It uses the inverse probability of exposure or treatment to balance the baseline features between the treated and the controls. (Chesnaye et al. 2022). First, we compute weights for the outcome model, which include weights for both mediator and treatment. We use the same strategy as conducted by VanderWeele and Tchetgen Tchetgen (2017), which sets all variables at time 1 will be baseline covariates. Therefore, the weights are (VanderWeele and Tchetgen Tchetgen 2017):

$$SW_{it}^{yt} = \frac{P(A_{it}=a_{it}|A_{it-1},M_{it-1})}{P(A_{it}=a_{it}|A_{it-1},M_{it-1},C,L_{it-1})} \tag{7}$$

$$SW_{it}^{ym} = \frac{P(M_{it}=m_{it}|M_{it-1},A_{it})}{P(M_{it}=m_{it}|M_{it-1},A_{it},C,L_{it-1})} \tag{8}$$

where $SW_{it}^{yt}$ are the weights for treatment in the outcome model for individual i for each time $t$. The numerator of $SW_{it}^{yt}$ represents the conditional probability of treatment at time $t$ conditioned on prior treatment and mediator at time $t-1$. The denominator of $SW_{it}^{yt}$ represents the conditional probability of treatment at time $t$ conditioned on prior mediator, treatment, and confounders at time $t-1$. The $SW_{it}^{ym}$ represent the weights for mediator in the outcome model. $SW_{it}^{yt} * SW_{it}^{ym}$ are used to estimate the coefficients in the outcome model. Similarly, the weights for the treatment in the mediation model are (VanderWeele and Tchetgen Tchetgen 2017):

$$SW_{it}^{mt} = \frac{P(A_{it}=a_{it}|A_{it-1})}{P(A_{it}=a_{it}|A_{it-1},M_{it-1},C,L_{it-1})} \tag{9}$$

However, in the IPTW, estimates could be impacted by extreme weights. Cole and Hernán (2008) propose using truncation to mitigate this issue, and following this proposal, we use 1% and 99% truncation to avoid extreme weights. The means for all stabilized weights in all models are around 1, and their SDs are less than 1. Further, no weights are excessively large; the maximum weight is 4.70, the minimum weight is 0.03 (see Appendix C for details).

## 4. Sensitivity Analysis

Causal mediation analysis based on interventional effects relies on a set of assumptions of which one of the most important is that there are no unobservable confounders between treatment-outcome, mediator-outcome, and treatment-mediator. Although we cannot directly test the assumption that there are no unmeasured confounders, we can nonetheless conduct a sensitivity analysis to test how sensitive our estimates are to the existence of such potential confounders. As noted by Lin et al. (2017), there is no formal sensitivity analysis to test these assumptions when using causal mediation analysis based on interventional effects. Therefore, we build on a recently proposed method to simulate the way these assumptions may impact our estimates (Bijlsma and Wilson 2020). This is achieved by assuming three unobservable confounders as follows.

**Step 1:** assume three equations for unmeasured treatment-mediator, treatment-outcome, and mediator-outcome confounders as functions of the variables that they confound, respectively.

$$V = \beta_0 + \beta_{1am}A + \beta_{2am}M + \varepsilon \quad (10)$$
$$U = \beta_0 + \beta_{1ay}A + \beta_{2ay}Y + \varepsilon \quad (11)$$
$$Z = \beta_0 + \beta_{1my}Y + \beta_{2my}M + \varepsilon \quad (12)$$

**Step 2:** manually set values for $\beta_{1am}, \beta_{2am}, \beta_{1ay}, \beta_{2ay}, \beta_{1my}, \beta_{2my}, \varepsilon \sim N(0, \delta^2)$, which specifies how large the unmeasured confounders are.

**Step 3:** predict unmeasured confounders according to the equations in Step 1.

**Step 4:** fit each of the regressions with unobservable confounders.

**Step 5:** check how the unmeasured confounders would impact the estimates.

**Step 6:** depending on the findings from Step 5, adjust the parameters at the Step 2.

In this setting, unobservable confounders include a set of potential confounders but no one specific confounder. There are also four further assumptions. First, we assume that the unmeasured confounders will not impact each other. Second, we assume that unmeasured confounders are only associated with the variables that they confound. Third, in this setting we assume that the unmeasured confounders will not be impacted by other variables controlled for in the models. In the current setting, this means assuming that there are no confounders that could impact treatment, mediator, and outcome simultaneously. Fourth, these unmeasured confounders are assumed to be pretreatment variables.

**Results**

**1. Descriptive Statistics**

Table 2 presents descriptive statistics for our analytical sample. For the time-varying covariates, we present snapshot-images for ages 20, 27, and 33 rather than for the full 17-year period. Notable patterns include that over the course of the period examined, individuals become increasingly concentrated to metropolitan counties, achieve higher levels of education and become homeowners, have more often had managerial experience, and fewer are unemployed. In addition, individuals' residential mobility decreases over time, and Their average scores on disadvantaged neighborhoods measure become smaller over time. Meanwhile, the share of individuals who receive social assistance also becomes smaller over time. The number of children aged 7-17 is quite high in the sample at age 20, with one possible reason for this being that individuals are more likely still to be living with their parents and siblings.

[Table 2 about here]

Figure 3 demonstrates how the use of social assistance varies over young adulthood, from 2001 to 2014. Figure 3A includes social assistance use by age. It shows that at age 20, 12.5 percent of our sample received social assistance. The share receiving social assistance during a given year then decreases over time, and at age 33, only about 3 percent are social assistance recipients. This is in line with official statistics and is probably due to the overall low unemployment rate in Sweden for this particular age group. Figure 3B includes cumulative social assistance recipiency over time, i.e., in how many years individuals had received social assistance by the time they reached age 33. The absolute majority, 76.6 percent, of our sample have never received social assistance. 7.6 percent received social assistance in 1 year and 6.6 percent received social assistance in 5 or more years. Figure 3C shows the interaction between cumulative social assistance recipiency and 14-year average disadvantaged neighborhood scores. Individuals with greater cumulative social assistance recipiency have a higher average exposure to disadvantaged neighborhoods. For instance, the 14-year average disadvantaged neighborhood score for individuals with 10 years of cumulative social assistance recipiency is about 0.43. However, the 14-year average disadvantaged neighborhood scores for the group with no cumulative social assistance recipiency is less than 0.35. This implies that the group with greater cumulative social assistance recipiency has been more exposed to less favorable neighborhoods. Figure 3D presents the relationship between cumulative social assistance recipiency and social assistance recipiency at age 34-36. The figure shows that those with higher levels of cumulative social assistance recipiency between age 20 and 33 also have a higher probability of receiving social assistance at age 34-36.

*[Figure 3 about here]*

**2. Social Assistance Dependency and the Mediating Effect of Neighborhoods**

All marginal structural models in this study use stabilized weights, which could improve the precision of the estimates. Table 3 presents the coefficients for each model. $\theta_1$ represents the cumulative total effect of past social assistance recipiency on future social assistance recipiency, and $\theta_2$ represents the cumulative total effect of neighborhood disadvantage.

Finally, $\beta_1$ represents the effect of past social assistance recipiency on neighborhood disadvantage. Using these coefficients, we can calculate the interventional effects. As noted above, we assumed that $\beta_0(t) = \beta_0$ and $\beta_1(t) = \beta_1$ are constant, and that for all times $t$, $a^*(t) = 0$ and $a(t) = 1$. We can then calculate the interventional direct effect $T\theta_1$ and the interventional indirect effect $T\beta_1\theta_2$ (VanderWeele and Tchetgen Tchetgen 2017), as presented in Figure 4.

*[Table 3 about here]*

Figure 4 presents the cumulative interventional effects with 95% confidence intervals, based on 500 bootstrapping replications. For the full model, the interventional direct effect (IDE) is 4.45 (95% CI: 4.31-4.61) on a risk ratio scale, which means that individuals who were exposed to cumulative social assistance recipiency, compared to those who were not exposed to cumulative social assistance recipiency, have 4.45 times the risk of social assistance recipiency at age 34-36 when their neighborhood disadvantage scores are randomly drawn from the distribution of individuals without cumulative social assistance recipiency. The interventional indirect effect (IIE) is 0.47 (95% CI: 0.34-0.60) on a risk ratio scale, which means that the difference in the risk of receiving social assistance at age 34-36 is 0.47 on a risk ratio scale when all individuals are cumulatively exposed to past social assistance recipiency and their neighborhood disadvantage scores are randomly drawn from the distribution of individuals with cumulative prior social assistance recipiency rather than drawn from those without cumulative prior social assistance recipiency, conditional on a set of covariates.

In order to calculate the proportion of the direct effect from cumulative social assistance recipiency that is mediated by the indirect effect from neighborhood disadvantage, we divide the indirect effect by the sum of the direct and indirect effects. This calculation shows that the proportion of the overall interventional effect that is mediated by the level of neighborhood disadvantage is 9.5%.

In sum, these results refute the null hypotheses of our hypotheses. Cumulative past social assistance recipiency may increase the risk for future social assistance recipiency. A moderate proportion of this relationship is mediated by the high levels of cumulative neighborhood disadvantage found among social assistance recipients.

*[Figure 4 about here]*

## 3. Sensitivity Analysis and Robustness Check

### 3.1 Robustness Check

To test the robustness of our results and interpretations, we have conducted a set of robustness checks, which are presented in Table 4. Model 0 in Table 4 is the original model. First, we re-estimate the models using a different individualized neighborhood size (Table 4, Model 1). In the main analyses, neighborhood characteristics are based on the 50 nearest neighbors, while in the robustness checks we instead base these characteristics on the 500 nearest neighbors. Our results show that the proportion of the overall interventional effect mediated by neighborhood disadvantage is 5.4%. One possible reason for this is that the k=500 individualized neighborhoods have a greater variance, which may lead to smaller estimates. This might in turn be due to the use of larger neighborhoods resulting in an underestimate of the concentration of disadvantage around disadvantaged individuals. Second, we re-estimate the models using a different sample, namely the cohort born in 1983 (Table 4, Model 2). The results are identical. Third, as noted above, our main analyses use annual data and define social assistance recipiency by whether an individual has received any social assistance during a given year, regardless of the magnitude or duration of this assistance. To ensure our results are not driven by small-scale recipients of social assistance, we re-define our measure of social assistance to only include individuals whose social assistance benefits account for more than 10% of their disposable income as social assistance recipients (Table 4, Model 3). Here the results shows that the proportion of the overall interventional effect mediated by neighborhood disadvantage is 7.8%, in line with our previous results

*[Table 4 about here]*

### 3.2 Sensitivity Analysis

In our sensitivity analyses, described in detail above, we constructed a set of different scenarios to test how sensitive our results are to unobserved confounders. For example, Figure 5A shows a naïve version of our sensitivity analyses under different scenarios, which indicates how unmeasured confounders between treatment and outcome impact the estimates. We include 5 scenarios by setting $\beta_{1ay}$ and $\beta_{2ay}$ to [0, 0.1, 0.3, 0.5, 1] simultaneously, and by fixing $\varepsilon \sim N(0,1)$, where these scenarios represent the size of the associations between unmeasured confounders and the treatment or outcome, respectively. Similarly, we also set two pairs ($\beta_{1am}, \beta_{2am}$) and ($\beta_{1my}, \beta_{2my}$) to [0, 0.1, 0.3, 0.5, 1] respectively. For instance, $\beta_{1ay}=0.5$ and $\beta_{2ay}=0.5$ mean that the size of the association between unmeasured confounders and the treatment is 0.5, and that the size of association between unmeasured confounders and the outcome is 0.5. In other words, these 5 scenarios include a spectrum of alternatives that range from the "weakest possible" unmeasured confounders to the "strongest possible" unmeasured confounders. Each scenario was simulated 500 times.

Figure 5A shows that unmeasured associations of greater than 0.3 could substantially bias the estimates of direct and indirect effects. Similar patterns are found in how unobservable treatment-outcome confounders could impact estimates (Figure 5B). However, Figure 5C demonstrates that the unmeasured confounders between treatment and mediator may not have much of an impact on estimates when these unmeasured associations are smaller than 1. However, as noted above, this is based on naïve assumptions regarding the relationships between unobservable confounders and other variables. If we keep these limitations in mind, the results from our sensitivity analyses provide approximate evidence that only relatively strong unmeasured confounders could substantially affect our estimates.

*[Figure 5 about here]*

**Discussion**

The research presented in this article improves our understanding about social assistance dependency in a number of ways. We study social assistance dependency by examining the cumulative effect of past social assistance recipiency on future assistance recipiency, rather than by examining how the duration of social assistance recipiency affects the risk of remaining reliant on social assistance. Using novel causal mediation analysis based on the counterfactual framework and Swedish register data, our findings suggest strong effects from past periods of social assistance on future social assistance recipiency, even though we allow for non-continuous recipiency spells.

Importantly, we provide new insights into the role of residential location for understanding social assistance dependency. Up until now, the neighborhood dimension has been almost completely neglected by research on social assistance dependency. This has limited our understanding of whether social assistance dependency is an individual-level process or is rather driven by an interplay with meso-level societal structures. In the present study, we have bridged this gap by investigating whether living in a disadvantaged neighborhood exacerbates social assistance dependency. More specifically, we have examined the mediating role played by cumulative exposure to disadvantaged neighborhoods on the effect of cumulative past social assistance recipiency on future recipiency. Our findings suggest that the overexposure to disadvantaged neighborhoods among social assistance recipients can explain 9.5 percent of the excess risk for future social assistance recipiency among individuals who have previously received social assistance. We interpret this as an effect size in the low to moderate range. However, as noted by Walters (2019), there are a number of situations that could lead to smaller mediation effects, i.e. loss of information over time, measurement error, and controlling for past treatments. Small mediation effects are common in observational studies and do not mean that the effects are unimportant (Walters 2019), since small mediation effects imply that these effects do exist even in the "worst" and "inauspicious" situations (Prentice and Miller 1992; Walters 2019). Based on these results, our models are relatively robust.

Even though our estimate of 9.5 percent means that 9.5 percent of social assistance dependency can be explained by adverse neighborhood composition, our results indicate that a large proportion of social assistance dependency is explained factors other than neighborhood context, such as individual-level attributes such as human capital deprivation. For example, scarring effects from unemployment in youth could harm individuals' human capital (Schemillen and Umkehrer 2017) and individuals' social networks might also be negatively affected by repeated spells on social assistance (Vandecasteele and Fasang 2020).

The rather modest mediating effect of exposure to disadvantaged neighborhoods found in this study might be understood through the lens of Swedish segregation patterns. In Sweden, the government introduced social-mix housing policies as early as the 1970s. These policies are more egalitarian and universal than those in many other western countries, including the US. The Swedish government's aim has been to create urban areas with decreased or no segregation by changing the composition of both disadvantaged and advantaged areas, whereas other countries' social-mix housing policies have often only targeted disadvantaged areas (Bergsten and Holmqvist 2013). In addition, the Swedish government has aimed to implement social-mix housing policies via city planning in connection with new construction rather than via planned demolition, which is common in other European countries, or via dispersal programs, such as those in the US (Holmqvist and Bergsten 2009). In general, social-mix housing policies focus on reducing economic segregation rather than ethnic segregation (Holmqvist and Bergsten 2009). For the purpose of our study, this may mean that individuals who rely on social assistance in Sweden are less overexposed to disadvantaged neighborhoods than similar individuals in other countries, which may explain the low or moderate mediating effect produced by such exposure. The research field would benefit from future studies studying this association in contexts with segregation patterns that differ from those found in Sweden, since we can expect country-level heterogeneity in the mediating role of disadvantaged neighborhoods.

There has been a critical change in housing policy in Sweden since the 1990s, from a welfare- to a market-based system. As noted by Sernhede, Thörn, and Thörn (2016), Stockholm is becoming a highly segregated city, with the market-based housing systems exacerbating residential segregation in the city (Andersson and Turner 2014; Bergsten and Holmqvist 2013). In 1990, all inner-city areas in Sweden had the target of achieving mixed neighborhoods, but central Stockholm instead became more homogenous as a result of tenure conversions in 2012 (Wimark, Andersson, and Malmberg 2020). However, the segregation patterns found in Gothenburg and Malmo differ from those of Stockholm (Bergsten and Holmqvist 2013; Wimark et al. 2020). In addition, ethnic segregation is found in both metropolitan and non-metropolitan areas (Malmberg et al. 2018). There is thus a diversity of segregation patterns in Sweden. As noted by Nieuwenhuis and Hooimeijer (2016), a high level of segregation might imply large neighborhood effects, and in future research we would therefore expect to find larger mediating effects for exposure to disadvantaged neighborhoods in more segregated areas, which also has implications for US neighborhood research. In general, levels of segregation are greater in the US than in western European and Nordic countries because the welfare systems in western Europe and the Nordic countries are more comprehensive than those of the US (Galster 2007; Nieuwenhuis and Hooimeijer 2016). Furthermore, segregation patterns are also diverse in the US, so we might expect research in the American context to find smaller neighborhood effects on welfare dependency in less segregated neighborhoods compared to those more segregated neighborhoods.

In addition, our results also further our understanding of cumulative disadvantage by suggesting a double-disadvantage process in which individuals who receive social assistance appear to experience both individual-level deprivation and neighborhood-level disadvantage, which jointly increase their likelihood of returning to social assistance recipiency. Cumulative and repeated social assistance recipiency constitutes a type of persistent poverty. Our results shows that the cumulative effects of micro-level and meso-level processes should be integrated to improve our understanding of persistent poverty. This could be viewed as a multilevel cumulative disadvantage framework. Both mechanisms of cumulative disadvantage (Lynn and Espy 2021) and theories of poverty (Brady 2019) emphasize the combination of micro-level and meso-level dynamics. Although we cannot in the present study identify the exact single mechanism through which the neighborhood effect operates at a lower abstraction level than the broader concept of neighborhood disadvantage, the results nonetheless show that meso-level dynamics do have a moderate impact on the micro-level association between past and future social assistance recipiency. Our results provide a coherent and reasonably parsimonious insight into how disadvantaged neighborhoods play a role in social assistance dependency, suggesting that this role involves both a spatial and temporal component. This also implies that policymakers' anti-poverty strategies should not only consider individual-level factors, but also take account of the way the social structure in which an individual is embedded may affect the chances of escaping poverty (Zhou and Liu 2019).

Importantly, this study also overcomes some major methodological challenges. First, the use of individualized neighborhoods to measure disadvantaged neighborhoods, rather than the administratively defined boundaries of geographical units, reduces the difficulties associated with the so-called modifiable areal unit problem (MAUP). Had we not used individualized neighborhoods, the neighbors of an individual living next to the border of an administrative area would have been assumed to be the same as the neighbors of an individual living on the opposite side of that area, which would risk introducing measurement error. Second, we use a temporal framework and causal mediation analysis to capture cumulative effects. Previous research on social assistance dependency has not been able to capture multiple recipiency spells or time-varying treatments with time-varying mediators. We use the mediated g-formula developed by VanderWeele and Tchetgen Tchetgen (2017) to mitigate these limitations. Third, as noted by VanderWeele et al. (2014), treatment-induced mediator-outcome confounders could bias the estimates of mediated effects. The assumption that there is no treatment-induced mediator-outcome confounders is often unrealistic. For this reason, we have used interventional effects to provide more precise estimates of mediated effects (VanderWeele et al. 2014). Importantly, the definition of interventional effects is based on population-level interventions, which means that these interventional effects are more policy-relevant (Demirer et al. 2021; Moreno-Betancur and Carlin 2018; Nguyen et al. 2021).

Although this study makes important contributions to theory and policy, it is not without limitations.

One possible limitation is our definitions for disadvantaged neighborhoods and social assistance recipiency. We base this on the share of neighbors with a low level of education, the share of low-income neighbors, the share of social assistance recipients, the share of unemployed neighbors, and the share of neighbors in low-skilled occupations and have constructed a "disadvantaged neighborhood score" using the k-nearest neighbor approach and PCA. This means that the index may not capture neighborhood disadvantage due to loss of information in the transformation. We measure social assistance recipiency annually, but the number receiving social assistance within a year may be small, which means that we may misestimate effect of cumulative social assistance recipiency.

Another important possible limitation is related to the fact that we cannot control for unobserved confounders using our chosen analytical strategy. If we have failed to control for some crucial confounder between treatments, mediators, and outcomes, our estimates will be biased. Psychosocial factors and biological factors might play unmeasured confounding roles at the individual level. DellaVigna et al. (2017) use the concept of reference dependence, which refers to the way individuals express preferences according to a status quo, as a means of explaining how individuals' preferences impact different probabilities of escaping from unemployment during different periods of unemployment insurance. Among those who are in the mid-term of their unemployment insurance benefits, by which time they have become used to having a low income from these benefits, individuals reduce their efforts to find a job, whereas the effort devoted to finding work is higher during the beginning of a spell of unemployment and directly prior to the expiry of unemployment insurance (DellaVigna et al. 2017) . In other words, the reference preference might constitute a potential psychological mechanism underlying the disincentive effect of welfare benefits (Schmieder and von Wachter 2016). However, given that social assistance does not have a fixed end point, this risk is smaller than it would have been if we had studied unemployment benefits. In addition, an individual's health status could impact both past and future social assistance recipiency and neighborhood selection. As noted by Bor, Cohen, and Galea (2017), poor health could lead to human capital loss and poverty, and poor health could also impede an individual's chances of leaving a disadvantaged neighborhood (Arcaya et al. 2014). There is also a risk that the link between past and future social assistance recipiency is impacted by the assessments of social workers, due to the presence of a wide margin of discretion (Hussénius 2021). Given that we do not know how social workers evaluate individuals based on their previous social assistance recipiency and/or residential context, this also constitutes a potential unobservable confounder.

To assess the extent to which our results might be driven by such potential confounders, we conducted a naïve sensitivity analysis. The results suggest that our results would remain robust even if we have omitted strong confounders. However, the naïve sensitivity analysis also has certain limitations. First, the artificial unobservable confounders are not perfectly predicted because we only assume simple relationships between unobservable confounders and treatment, mediator, and outcome. Second, we did not assume the existence of a treatment-mediator interaction because we assumed the additive process in social assistance dependency or other complicated relationships between different variables and unobservable confounders.

With these caveats in mind, the results presented in this paper provide important evidence of the low to moderate mediating role of neighborhood in social assistance dependency. In sum, our results suggest that in order to reduce social assistance dependency, we need to consider both individual-level situations, and individuals' living conditions. Although cumulative neighborhood disadvantage only accounts for 9.5% of overall cumulative social assistance dependency, the effect is robust. The development of a formal sensitivity analysis and testing the role and heterogeneity of disadvantaged neighborhoods in relation to poverty dynamics in different contexts both constitute key issues for future research.

**REFERENCES**

Amcoff, J., Östh, J., Niedomysl, T., Engkvist, R., Moberg, U. 2014. "Segregation i Stockholmsregionen: Kartläggning med EquiPop [Segregation in the Stockholm region: Mapping with EquiPop]". In Befolkningsprognos 2014-2023/45; *Demografisk rapport* 2014:09 (2014). Stockholm County Council.

Aizer, Anna, Hilary W. Hoynes, and Adriana Lleras-Muney. 2022. "Children and the US Social Safety Net: Balancing Disincentives for Adults and Benefits for Children." *National Bureau of Economic Research Working Paper Series* No. 29754. doi: 10.3386/w29754.

Andersson, Roger, and Lena Magnusson Turner. 2014. "Segregation, Gentrification, and Residualisation: From Public Housing to Market-Driven Housing Allocation in Inner City Stockholm." *International Journal of Housing Policy* 14(1):3–29. doi: 10.1080/14616718.2013.872949.

Arcaya, Mariana C., S. V Subramanian, Jean E. Rhodes, and Mary C. Waters. 2014. "Role of Health in Predicting Moves to Poor Neighborhoods among Hurricane Katrina Survivors." *Proceedings of the National Academy of Sciences* 111(46):16246. doi: 10.1073/pnas.1416950111.

Åslund, Olof, and Peter Fredriksson. 2009. "Peer Effects in Welfare Dependence: Quasi-Experimental Evidence." *The Journal of Human Resources* 44(3):798–825. doi: 10.3368/jhr.44.3.798

Bäckman, Olof, and Åke Bergmark. 2011. "Escaping Welfare? Social Assistance Dynamics in Sweden." *Journal of European Social Policy* 21(5):486–500. doi: 10.1177/0958928711418855.

Baird, Sarah, David McKenzie, and Berk Özler. 2018. "The Effects of Cash Transfers on Adult Labor Market Outcomes." *IZA Journal of Development and Migration* 8(1):22. doi: 10.1186/s40176-018-0131-9.

Banerjee, Abhijit V, Rema Hanna, Gabriel E. Kreindler, and Benjamin A. Olken. 2017. "Debunking the Stereotype of the Lazy Welfare Recipient: Evidence from Cash Transfer Programs." *World Bank Research Observer* 32(2):155–84. doi: 10.1093/wbro/lkx002.

Ben-Shlomo, Yoav, Gita Mishra, and Diana Kuh. 2014. "Life Course Epidemiology" Pp. 1521–49 in *Handbook of Epidemiology*, edited by W. Ahrens and I. Pigeot. New York, NY: Springer New York. https://doi.org/10.1007/978-0-387-09834-0_56

Benard, Stephen, and Robb Willer. 2007. "A Wealth and Status-Based Model of Residential Segregation." *The Journal of Mathematical Sociology* 31(2):149–74. doi:

10.1080/00222500601188486.

Berg, Noora. 2017. "Accumulation of disadvantage from adolescence to midlife: A 26-year follow-up study of 16-year old adolescents." *Dissertationes Scholae Doctoralis Ad Sanitatem Investigandam Universitatis Helsinkiensis.* Retrieved from DiVA. ISSN 2342-317X ; 8/2017 1

Bergman, Peter, Raj Chetty, Stefanie DeLuca, Nathaniel Hendren, Lawrence F. Katz, and Christopher Palmer. 2019. "Creating Moves to Opportunity: Experimental Evidence on Barriers to Neighborhood Choice." *National Bureau of Economic Research Working Paper Series* No. 26164. doi: 10.3386/w26164.

Bergsten, Zara, and Emma Holmqvist. 2013. "Possibilities of Building a Mixed City – Evidence from Swedish Cities." *International Journal of Housing Policy* 13(3):288–311. doi: 10.1080/14616718.2013.809211.

Bijlsma, Maarten J., and Ben Wilson. 2020. "Modelling the Socio-Economic Determinants of Fertility: A Mediation Analysis Using the Parametric g-Formula." *Journal of the Royal Statistical Society: Series A (Statistics in Society)* 183(2):493–513. doi: https://doi.org/10.1111/rssa.12520.

Birkelund, Gunn Elisabeth, Kristian Heggebø, and Jon Rogstad. 2017. "Additive or Multiplicative Disadvantage? The Scarring Effects of Unemployment for Ethnic Minorities." *European Sociological Review* 33(1):17–29. doi: 10.1093/esr/jcw030.

Birnbaum, Simon. 2010. "Radical Liberalism, Rawls and the Welfare State: Justifying the Politics of Basic Income." *Critical Review of International Social and Political Philosophy* 13(4):495–516. doi: 10.1080/09692290.2010.517968.

Blank, Rebecca M. 1989. "Analyzing the Length of Welfare Spells." *Journal of Public Economics* 39(3):245–73. doi: https://doi.org/10.1016/0047-2727(89)90029-7.

Bolt, Gideon, and Ronald van Kempen. 2013. "Introduction Special Issue: Mixing

Neighbourhoods: Success or Failure?" *Cities* 35:391–96. doi: https://doi.org/10.1016/j.cities.2013.04.006.

Bor, Jacob, Gregory H. Cohen, and Sandro Galea. 2017. "Population Health in an Era of Rising Income Inequality: USA, 1980–2015." *The Lancet* 389(10077):1475–90. doi: https://doi.org/10.1016/S0140-6736(17)30571-8.

Boschman, Sanne, Ineke Maas, J. Cok Vrooman, and Marcus H. Kristiansen. 2021. "From Social Assistance to Self-Sufficiency: Low Income Work as a Stepping Stone." *European Sociological Review*. doi: 10.1093/esr/jcab003.

Brady, David. 2019. "Theories of the Causes of Poverty." *Annual Review of Sociology* 45(1):155–75. doi: 10.1146/annurev-soc-073018-022550.

Browning, Christopher R., Catherine A. Calder, Lauren J. Krivo, Anna L. Smith, and Bethany Boettner. 2017. "Socioeconomic Segregation of Activity Spaces in Urban Neighborhoods: Does Shared Residence Mean Shared Routines?" *RSF: The Russell Sage Foundation Journal of the Social Sciences* 3(2):210. doi: 10.7758/RSF.2017.3.2.09.

Bunel, Mathieu, Yannick L'Horty, and Pascale Petit. 2015. "Discrimination Based on Place of Residence and Access to Employment." *Urban Studies* 53(2):267–86. doi: 10.1177/0042098014563470.

Camina, M. M., and M. J. Wood. 2009. "Parallel Lives: Towards a Greater Understanding of What Mixed Communities Can Offer." *Urban Studies* 46(2):459–80. doi: 10.1177/0042098008099363.

Carlsson, Magnus, Abdulaziz Abrar Reshid, and Dan-Olof Rooth. 2018. "Neighborhood Signaling Effects, Commuting Time, and Employment." *International Journal of Manpower* 39(4):534–49. doi: 10.1108/IJM-09-2017-0234.

Carpentier, Sarah, Karel Neels, and Karel Van den Bosch. 2017. "Exit from and Re-Entry

into Social Assistance Benefit in Belgium among People with Migration Background and the Native-Born." *International Journal of Social Welfare* 26(4):366–83. doi: https://doi.org/10.1111/ijsw.12270.

Casciano, Rebecca, and Douglas S. Massey. 2008. "Neighborhoods, Employment, and Welfare Use: Assessing the Influence of Neighborhood Socioeconomic Composition." *Social Science Research* 37(2):544–58. doi: https://doi.org/10.1016/j.ssresearch.2007.08.008.

Chen, Wansu, Lei Qian, Jiaxiao Shi, and Meredith Franklin. 2018. "Comparing Performance between Log-Binomial and Robust Poisson Regression Models for Estimating Risk Ratios under Model Misspecification." *BMC Medical Research Methodology* 18(1):63. doi: 10.1186/s12874-018-0519-5.

Chesnaye, Nicholas C., Vianda S. Stel, Giovanni Tripepi, Friedo W. Dekker, Edouard L. Fu, Carmine Zoccali, and Kitty J. Jager. 2022. "An Introduction to Inverse Probability of Treatment Weighting in Observational Research." *Clinical Kidney Journal* 15(1):14–20. doi: 10.1093/ckj/sfab158.

Chetty, Raj, Nathaniel Hendren, and Lawrence F. Katz. 2016. "The Effects of Exposure to Better Neighborhoods on Children: New Evidence from the Moving to Opportunity Experiment." *American Economic Review* 106(4):855–902.

Chyn, Eric. 2018. "Moved to Opportunity: The Long-Run Effects of Public Housing Demolition on Children." *American Economic Review* 108(10):3028–56. doi: 10.1257/aer.20161352.

Chyn, Eric, and Lawrence F. Katz. 2021. "Neighborhoods Matter: Assessing the Evidence for Place Effects." *Journal of Economic Perspectives* 35(4):197–222. doi: 10.1257/jep.35.4.197.

Cole, Stephen R., and Miguel A. Hernán. 2008. "Constructing Inverse Probability Weights

for Marginal Structural Models." *American Journal of Epidemiology* 168(6):656–64. doi: 10.1093/aje/kwn164.

Collins, William J., and Marianne H. Wanamaker. 2014. "Selection and Economic Gains in the Great Migration of African Americans: New Evidence from Linked Census Data." *American Economic Journal: Applied Economics* 6(1):220–52. doi: 10.1257/app.6.1.220.

Contini, Dalit, and Nicola Negri. 2007. "Would Declining Exit Rates from Welfare Provide Evidence of Welfare Dependence in Homogeneous Environments?" *European Sociological Review* 23(1):21–33. doi: 10.1093/esr/jcl017.

Cooke, Martin. 2009. "A Welfare Trap? The Duration and Dynamics of Social Assistance Use among Lone Mothers in Canada." *Canadian Review of Sociology/Revue Canadienne de Sociologie* 46(3):179–206. doi: https://doi.org/10.1111/j.1755-618X.2009.01211.x.

Custers, Gijs. 2019. "Neighbourhood Ties and Employment: A Test of Different Hypotheses across Neighbourhoods." *Housing Studies* 34(7):1212–34. doi: 10.1080/02673037.2018.1527020.

Dahl, Gordon B., Andreas Ravndal Kostøl, and Magne Mogstad. 2014. "Family Welfare Cultures." *The Quarterly Journal of Economics* 129(4):1711–52. doi: 10.1093/qje/qju019.

DellaVigna, Stefano, Attila Lindner, Balázs Reizer, and Johannes F. Schmieder. 2017. "Reference-Dependent Job Search: Evidence from Hungary." *The Quarterly Journal of Economics* 132(4):1969–2018. doi: 10.1093/qje/qjx015.

Demirer, Ibrahim, Michael Kühhirt, Ute Karbach, and Holger Pfaff. 2021. "Does Positive Affect Mediate the Association of Multimorbidity on Depressive Symptoms?" *Aging & Mental Health* 1–12. doi: 10.1080/13607863.2020.1870209.

Deryugina, Tatyana, and David Molitor. 2020. "Does When You Die Depend on Where You Live? Evidence from Hurricane Katrina." *American Economic Review* 110(11):3602–33. doi: 10.1257/aer.20181026.

Desmond, Matthew, and Nathan Wilmers. 2019. "Do the Poor Pay More for Housing? Exploitation, Profit, and Risk in Rental Markets." *American Journal of Sociology* 124(4):1090–1124. doi: 10.1086/701697.

DiPrete, Thomas A., and Gregory M. Eirich. 2006. "Cumulative Advantage as a Mechanism for Inequality: A Review of Theoretical and Empirical Developments." *Annual Review of Sociology* 32(1):271–97. doi: 10.1146/annurev.soc.32.061604.123127.

Fjellborg, Andreas Alm. 2021. "Residential Mobility and Spatial Sorting in Stockholm 1990-2014: The Changing Importance of Housing Tenure and Income." *International Journal of Housing Policy* 1–21. doi: 10.1080/19491247.2021.1893117.

Galster, George. 2007. "Should Policy Makers Strive for Neighborhood Social Mix? An Analysis of the Western European Evidence Base." *Housing Studies* 22(4):523–45. doi: 10.1080/02673030701387630.

Galster, George C. 2012. "The Mechanism(s) of Neighbourhood Effects: Theory, Evidence, and Policy Implications" Pp. 23–56 in *Neighbourhood Effects Research: New Perspectives*, edited by M. van Ham, D. Manley, N. Bailey, L. Simpson, and D. Maclennan. Dordrecht: Springer Netherlands. https://doi.org/10.1007/978-94-007-2309-2_2

Galster, George, Dave E. Marcotte, Marv Mandell, Hal Wolman, and Nancy Augustine. 2007. "The Influence of Neighborhood Poverty During Childhood on Fertility, Education, and Earnings Outcomes." *Housing Studies* 22(5):723–51. doi: 10.1080/02673030701474669.

Grander, Martin. 2017. "New Public Housing: A Selective Model Disguised as Universal?

Implications of the Market Adaptation of Swedish Public Housing." *International Journal of Housing Policy* 17(3):335–52. doi: 10.1080/19491247.2016.1265266.

Grander, Martin. 2021. "The Inbetweeners of the Housing Markets – Young Adults Facing Housing Inequality in Malmö, Sweden." *Housing Studies* 1–18. doi: 10.1080/02673037.2021.1893278.

Gustafsson, Björn, Katarina Katz, and Torun Osterberg. 2019. "Social Assistance Receipt Among Young Adults Who Grow Up in Different Neighborhoods of Metropolitan Sweden." *Poverty & Public Policy* 11(4):302–24. doi: https://doi.org/10.1002/pop4.264.

Harjunen, Oskari; Saarimaa, Tuukka; Tukiainen, Janne .2021. "Love Thy (Elected) Neighbor? Residential Segregation, Political Representation and Local Public Goods." Discussion paper, No. 138, Aboa Centre for Economics (ACE), Turku http://dx.doi.org/10.2139/ssrn.3765518

Harding, David J. 2010. *Living the Drama: Community, Conflict, and Culture among Inner-City Boys*. University of Chicago Press.

Hedman, Lina. 2014. "Compositional or Contextual Effects? Neighbourhoods and Teenage Parenthood in Stockholm, Sweden." *KZfSS Kölner Zeitschrift Für Soziologie Und Sozialpsychologie* 66(1):67–90. doi: 10.1007/s11577-014-0270-9.

Hedstrom, Peter. 2005. *Dissecting the Social: On the Principles of Analytical Sociology*. Cambridge University Press.

Hennerdal, Pontus, and Michael Meinild Nielsen. 2017. "A Multiscalar Approach for Identifying Clusters and Segregation Patterns That Avoids the Modifiable Areal Unit Problem." *Annals of the American Association of Geographers* 107(3):555–74. doi: 10.1080/24694452.2016.1261685.

Holmqvist, Emma, and Zara Bergsten. 2009. "Swedish Social Mix Policy: A General Policy without an Explicit Ethnic Focus." *Journal of Housing and the Built Environment*

24(4):477. doi: 10.1007/s10901-009-9162-0.

Hümbelin, Oliver, and Tobias Fritschi. 2018. "Pathways into and out of the Labor Market After Receiving Social Benefits: Cumulative Disadvantage or Life Course Risk?" *The Sociological Quarterly* 59(4):627–54. doi: 10.1080/00380253.2018.1489207.

Hussénius, Klara. 2021. "Intersectional Patterns of Social Assistance Eligibility in Sweden." *Nordic Social Work Research* 11(1):19–33. doi: 10.1080/2156857X.2019.1601636.

Ilmakunnas, Ilari. 2018. "Risk and Vulnerability in Social Assistance Receipt of Young Adults in Finland." *International Journal of Social Welfare* 27(1):5–16. doi: https://doi.org/10.1111/ijsw.12273.

Ilmakunnas, Ilari, and Pasi Moisio. 2019. "Social Assistance Trajectories among Young Adults in Finland: What Are the Determinants of Welfare Dependency?" *Social Policy & Administration* 53(5):693–708. doi: https://doi.org/10.1111/spol.12413.

Jivraj, Stephen, Emily T. Murray, Paul Norman, and Owen Nicholas. 2020. "The Impact of Life Course Exposures to Neighbourhood Deprivation on Health and Well-Being: A Review of the Long-Term Neighbourhood Effects Literature." *European Journal of Public Health* 30(5):922–28. doi: 10.1093/eurpub/ckz153.

Johnson, Pamela Jo, J. Michael Oakes, and Douglas L. Anderton. 2008. "Neighborhood Poverty and American Indian Infant Death: Are The Effects Identifiable?" *Annals of Epidemiology* 18(7):552–59. doi: https://doi.org/10.1016/j.annepidem.2008.02.007.

Königs, Sebastian. 2018. "Micro-Level Dynamics of Social Assistance Receipt: Evidence from Four European Countries." *International Journal of Social Welfare* 27(2):146–56. doi: https://doi.org/10.1111/ijsw.12279.

Korpi, Walter, and Joakim Palme. 1998. "The Paradox of Redistribution and Strategies of Equality: Welfare State Institutions, Inequality, and Poverty in the Western Countries." *American Sociological Review* 63(5):661–87. doi: 10.2307/2657333.

Kwan, Mei-Po. 2018. "The Limits of the Neighborhood Effect: Contextual Uncertainties in Geographic, Environmental Health, and Social Science Research." *Annals of the American Association of Geographers* 108(6):1482–90. doi: 10.1080/24694452.2018.1453777.

Leventhal, Tama, and Veronique Dupéré. 2019. "Neighborhood Effects on Children's Development in Experimental and Nonexperimental Research." *Annual Review of Developmental Psychology* 1(1):149–76. doi: 10.1146/annurev-devpsych-121318-085221.

Lin, Sheng-Hsuan, Jessica Young, Roger Logan, Eric J. Tchetgen Tchetgen, and Tyler J. VanderWeele. 2017. "Parametric Mediational G-Formula Approach to Mediation Analysis with Time-Varying Exposures, Mediators, and Confounders." *Epidemiology* 28(2): 266–274. doi: 10.1097/EDE.0000000000000609

Lorentzen, Thomas, Anna Angelin, Espen Dahl, Timo Kauppinen, Pasi Moisio, and Tapio Salonen. 2014. "Unemployment and Economic Security for Young Adults in Finland, Norway and Sweden: From Unemployment Protection to Poverty Relief." *International Journal of Social Welfare* 23(1):41–51. doi: 10.1111/ijsw.12006.

Ludwig, Jens, Lisa Sanbonmatsu, Lisa Gennetian, Emma Adam, Greg J. Duncan, Lawrence F. Katz, Ronald C. Kessler, Jeffrey R. Kling, Stacy Tessler Lindau, Robert C. Whitaker, and Thomas W. McDade. 2011. "Neighborhoods, Obesity, and Diabetes — A Randomized Social Experiment." *New England Journal of Medicine* 365(16):1509–19. doi: 10.1056/NEJMsa1103216.

Lundberg, Ian, Rebecca Johnson, and Brandon M. Stewart. 2021. "What Is Your Estimand? Defining the Target Quantity Connects Statistical Evidence to Theory." *American Sociological Review* 86(3):532–65. doi: 10.1177/00031224211004187.

Lynn, Freda B., and Hannah W. Espy. 2021. "Cumulative Advantage." Pp.286-307 in

*Research Handbook on Analytical Sociology*, edited by Gianluca Manzo. Cheltenham, UK: Edward Elgar Publishing. doi: https://doi.org/10.4337/9781789906851.00023

Malmberg, Bo, and Eva Andersson. 2016. "Analysing Segregation with Individualized Neighbourhoods Defined by Population Size." (August 2014):1990–2016. doi: 10.1332/policypress/9781447301356.003.0007.

Malmberg, Bo, Eva K. Andersson, Michael M. Nielsen, and Karen Haandrikman. 2018. "Residential Segregation of European and Non-European Migrants in Sweden: 1990–2012." *European Journal of Population* 34(2):169–93. doi: 10.1007/s10680-018-9478-0.

Markussen, Simen, and Knut Røed. 2016. "Leaving Poverty Behind? The Effects of Generous Income Support Paired with Activation." *American Economic Journal: Economic Policy* 8(1):180–211. doi: 10.1257/pol.20140334.

Marttila, Anneli, Margaret Whitehead, Krysia Canvin, and Bo Burström. 2010. "Controlled and Dependent: Experiences of Living on Social Assistance in Sweden." *International Journal of Social Welfare* 19(2):142–51. doi: https://doi.org/10.1111/j.1468-2397.2009.00638.x.

Meyer, Bruce D. 1990. "Unemployment Insurance and Unemployment Spells." *Econometrica* 58(4):757–82. doi: 10.2307/2938349.

Mood, Carina. 2010. "Neighborhood Social Influence and Welfare Receipt in Sweden: A Panel Data Analysis." *Social Forces* 88(3):1331–56. doi: 10.1353/sof.0.0304.

Mood, Carina. 2013. "Social Assistance Dynamics in Sweden: Duration Dependence and Heterogeneity." *Social Science Research* 42(1):120–39. doi: https://doi.org/10.1016/j.ssresearch.2012.07.005.

Moreno-Betancur, Margarita, and John B. Carlin. 2018. "Understanding Interventional Effects: A More Natural Approach to Mediation Analysis?" *Epidemiology* 29(5):614-

617. doi: 10.1097/EDE.0000000000000866.

Nguyen, Trang Quynh, Ian Schmid, and Elizabeth A. Stuart. 2021. "Clarifying Causal Mediation Analysis for the Applied Researcher: Defining Effects Based on What We Want to Learn." *Psychological Methods* 26(2):255–71. doi: 10.1037/met0000299

Nieuwenhuis, Jaap. 2016. "Publication Bias in the Neighbourhood Effects Literature." *Geoforum* 70:89–92. doi: https://doi.org/10.1016/j.geoforum.2016.02.017.

Nieuwenhuis, Jaap, and Pieter Hooimeijer. 2016. "The Association between Neighbourhoods and Educational Achievement, a Systematic Review and Meta-Analysis." *Journal of Housing and the Built Environment* 31(2):321–47. doi: 10.1007/s10901-015-9460-7.

Östh, John, William A. V Clark, and Bo Malmberg. 2015. "Measuring the Scale of Segregation Using K-Nearest Neighbor Aggregates." *Geographical Analysis* 47(1):34–49. doi: 10.1111/gean.12053.

Pearl, Judea. 2014. "Interpretation and Identification of Causal Mediation." *Psychological Methods* 19(4):459–81. doi: 10.1037/a0036434.

Pinkster, Fenne M. 2014. "Neighbourhood Effects as Indirect Effects: Evidence from a Dutch Case Study on the Significance of Neighbourhood for Employment Trajectories." *International Journal of Urban and Regional Research* 38(6):2042–59. doi: https://doi.org/10.1111/j.1468-2427.2012.01197.x.

Pissarides, Christopher A. 1992. "Loss of Skill During Unemployment and the Persistence of Employment Shocks." *The Quarterly Journal of Economics* 107(4):1371–91. doi: 10.2307/2118392.

Prentice, Deborah A., and Dale T. Miller. 1992. "When Small Effects Are Impressive." *Psychological Bulletin* 112(1):160–64. doi: 10.1037/0033-2909.112.1.160.

Rawls, John. 1971. *A Theory of Justice*. Harvard University Press.

Reardon, Sean F., and Kendra Bischoff. 2011. "Income Inequality and Income Segregation."

*American Journal of Sociology* 116(4):1092–1153. doi: 10.1086/657114.

Robins, James M., Miguel Ángel Hernán, and Babette Brumback. 2000. "Marginal Structural Models and Causal Inference in Epidemiology." *Epidemiology* 11(5):550–60. doi: 10.1097/00001648-200009000-00011.

Sampson, Robert J. 2011. "Neighborhood Effects, Causal Mechanisms and the Social Structure of the City." Pp. 227–49 in *Analytical Sociology and Social Mechanisms*, edited by P. Demeulenaere. Cambridge: Cambridge University Press.

Sampson, Robert J., and Patrick Sharkey. 2008. "Neighborhood Selection and the Social Reproduction of Concentrated Racial Inequality." *Demography* 45(1):1–29. doi: 10.1353/dem.2008.0012.

Schachner, Jared N., and Robert J. Sampson. 2020. "Skill-Based Contextual Sorting: How Parental Cognition and Residential Mobility Produce Unequal Environments for Children." *Demography* 57(2):675–703. doi: 10.1007/s13524-020-00866-8.

Schmieder, Johannes F., and Till von Wachter. 2016. "The Effects of Unemployment Insurance Benefits: New Evidence and Interpretation." *Annual Review of Economics* 8(1):547–81. doi: 10.1146/annurev-economics-080614-115758.

Schmillen, Achim, and Matthias Umkehrer. 2017. "The Scars of Youth: Effects of Early-Career Unemployment on Future Unemployment Experience." *International Labour Review* 156(3–4):465–94. doi: https://doi.org/10.1111/ilr.12079.

Schootman, Mario, Elena M. Andresen, Fredric D. Wolinsky, Theodore K. Malmstrom, J. Philip Miller, and Douglas K. Miller. 2007. "Neighbourhood Environment and the Incidence of Depressive Symptoms among Middle-Aged African Americans." *Journal of Epidemiology and Community Health* 61(6):527 LP – 532. doi: 10.1136/jech.2006.050088.

Sernhede, Ove, Catharina Thörn, and Håkan Thörn. 2016. "The Stockholm Uprising in

Context: Urban Social Movements in the Rise and Demise of the Swedish Welfare-State City" Pp. 149–73 in *Urban Uprisings: Challenging Neoliberal Urbanism in Europe*, edited by M. Mayer, C. Thörn, and H. Thörn. London: Palgrave Macmillan UK. doi: https://doi.org/10.1057/978-1-137-50509-5_5

Stenberg, Sten-Åke. 2000. "Inheritance of Welfare Recipiency: An Intergenerational Study of Social Assistance Recipiency in Postwar Sweden." *Journal of Marriage and Family* 62(1):228–39. doi: https://doi.org/10.1111/j.1741-3737.2000.00228.x

Stranz, Hugo, and Stefan Wiklund. 2012. "Risk Factors of Long-Term Social Assistance Recipiency among Lone Mothers. The Case of Sweden." *European Journal of Social Work* 15(4):514–31. doi: 10.1080/13691457.2012.702312.

Tseng, Yi-Ping, Ha Vu, and Roger Wilkins. 2008. "Dynamic Properties of Income Support Receipt in Australia." *Australian Economic Review* 41(1):32–55. doi: https://doi.org/10.1111/j.1467-8462.2008.00474.x.

Van der Klaauw, Bas, and Jan C. van Ours. 2003. "From Welfare to Work: Does the Neighborhood Matter?" *Journal of Public Economics* 87(5):957–85. doi: https://doi.org/10.1016/S0047-2727(01)00133-5.

Vandecasteele, Leen, and Anette Eva Fasang. 2020. "Neighbourhoods, Networks and Unemployment: The Role of Neighbourhood Disadvantage and Local Networks in Taking up Work." *Urban Studies* 58(4):696–714. doi: 10.1177/0042098020925374.

VanderWeele, Tyler J. 2019. "Principles of Confounder Selection." *European Journal of Epidemiology* 34(3):211–19. doi: 10.1007/s10654-019-00494-6.

VanderWeele, Tyler J., and Eric J. Tchetgen Tchetgen. 2017. "Mediation Analysis with Time Varying Exposures and Mediators." *Journal of the Royal Statistical Society: Series B (Statistical Methodology)* 79(3):917–38. doi: https://doi.org/10.1111/rssb.12194.

VanderWeele, Tyler J., Linda Valeri, and Cande V Ananth. 2019. "Counterpoint: Mediation

Formulas With Binary Mediators and Outcomes and the 'Rare Outcome Assumption.'" *American Journal of Epidemiology* 188(7):1204–5. doi: 10.1093/aje/kwy281.

VanderWeele, Tyler J., and Stijn Vansteelandt. 2010. "Odds Ratios for Mediation Analysis for a Dichotomous Outcome." *American Journal of Epidemiology* 172(12):1339–48. doi: 10.1093/aje/kwq332.

VanderWeele, Tyler J., Stijn Vansteelandt, and James M. Robins. 2014. "Effect Decomposition in the Presence of an Exposure-Induced Mediator-Outcome Confounder." *Epidemiology* 25(2):300–306. doi: 10.1097/EDE.0000000000000034

Walters, Glenn D. 2019. "Why Are Mediation Effects so Small?" *International Journal of Social Research Methodology* 22(2):219–32. doi: 10.1080/13645579.2018.1517232.

Wilson, William Julius. 2012. *The Truly Disadvantaged: The Inner City, the Underclass, and Public Policy*. University of Chicago Press.

Wimark, Thomas, Eva K. Andersson, and Bo Malmberg. 2020. "Tenure Type Landscapes and Housing Market Change: A Geographical Perspective on Neo-Liberalization in Sweden." *Housing Studies* 35(2):214–37. doi: 10.1080/02673037.2019.1595535.

Wodtke, Geoffrey T., David J. Harding, and Felix Elwert. 2011. "Neighborhood Effects in Temporal Perspective: The Impact of Long-Term Exposure to Concentrated Disadvantage on High School Graduation." *American Sociological Review* 76(5):713–36. doi: 10.1177/0003122411420816.

Wodtke, Geoffrey T., Sagi Ramaj, and Jared Schachner. 2022. "Toxic Neighborhoods: The Effects of Concentrated Poverty and Environmental Lead Contamination on Early Childhood Development." *Demography* 59(4):1275–98. doi: 10.1215/00703370-10047481.

Wodtke, Geoffrey T., and Xiang Zhou. 2020. "Effect Decomposition in the Presence of Treatment-Induced Confounding: A Regression-with-Residuals Approach."

*Epidemiology* (435):369–75. doi: 10.1097/EDE.0000000000001168.

Zhou, Yang, and Yansui Liu. 2019. "The Geography of Poverty: Review and Research Prospects." *Journal of Rural Studies* (January). doi: 10.1016/j.jrurstud.2019.01.008.

Zou, Guangyong. 2004. "A Modified Poisson Regression Approach to Prospective Studies with Binary Data." *American Journal of Epidemiology* 159(7):702–6. doi: 10.1093/aje/kwh090.

**Appendixes**

**Appendix A. Defining Individualized Neighborhoods**

The method was developed by Östh et al. (2015). Using Swedish register data and geographical data, we can obtain locations based on 100 by 100 meter squares in which individuals live. The first step is then to calculate the total population that lives in the same square. The second step involves setting a k-neighbors threshold, which specifies how many neighbors will be included in an individualized neighborhood. In our setting, we have chosen k=50 but have also examined the effects of using different k-values in our sensitivity analysis. The third step is then to calculate the distance between different squares and aggregate neighborhoods in relation to individuals. For example, if k is equal to 50 for individual i, then the algorithm will calculate the total population of individual i's square and of other squares around individual i based on the shortest Euclidean distances between these squares. If the total population of squares for individual i is equal to 50, the algorithm will stop. The final step involves choosing the main variables that will be used to specify individualized neighborhoods. In our setting, we use the share of less-well educated, the share of low-income inhabitants, the share of social assistance recipients, the share who are unemployed and the share of neighbors with low-skilled occupations. It should be noted that the total populations of squares for individual i are not always equal to 50 but could and often be larger than 50 because of different populations within different squares. Finally, we use five variables: (1) the share of neighbors with only compulsory schooling, (2) the share of neighbors in the lowest income decile, (3) the share of neighbors who are social assistance recipients, (4) the share of neighbors who have been registered as unemployed at any time during the year, and (5) the share of neighbors who are employed in a low-skilled occupation to specify the final disadvantaged neighborhood scores variable using principal component analysis (PCA). In the results of the PCA, the proportion of total variance explained for each year is larger than 0.6, which constitutes an acceptable level for the construction of a single index.

## Appendix B. Motivations of Confounder Selection

**Motivations of Confounder Selection**

| Measurements | Definitions | Motivations |
|---|---|---|
| **Time-varying confounders** | | |
| Marital status | The situation indicates if an individual is single, married, divorced, or widowed. | Demographical factor |
| Living in metropolitan county | Individual lives in a metropolitan area in Sweden | Living in a metropolitan county may impact individuals' job opportunities |
| Education | Individual's level of education | Demographic factor |
| Residential mobility within a year | Individual's number of residential moves within a year | The higher the frequency of residential mobility, the less likely an individual is to make connections with others |
| Number of children aged 7-17 | | Demographic factor |
| Number of children aged 4-6 | | Demographic factor |
| Number of children aged 0-3 | | Demographic factor |
| Unemployment status | Individual has been registered as jobseeker at public employment agency | Demographic factor |
| Housing allowance | Individual has received housing-relevant benefits from government | Housing allowance may impact residential choice and social assistance |
| Parental benefits | Individual has received parental benefits from government | Housing allowance may impact social assistance use |
| Number of companies within the local labor market | | Local labor market could impact job opportunities |
| Size of working-age population within the local labor market | Working age specified as population aged 18-64 | Local labor market could impact job opportunities |
| Experience of working in a skilled occupation | Individual has experience of occupation requiring university-level qualification or equivalent | Better occupations may impact social assistance use |
| Living in non-rental housing | living in an owned house or apartment (although we do not know if whether the individual is the actual owner of the property), | Impacts where individuals live and social assistance use |
| Disposable income (log) excluding social assistance and parental benefit and housing allowance | | Impacts both social assistance recipiency and where individuals live |

**Time-invariant Confounders**

| | | |
|---|---|---|
| Gender | | Demographic factor |
| Migration Background | | Demographic factor |

**Appendix C. Truncated IPTWs**

**Properties of Calculated Weights**

| **Truncated IPWs** | Mean | SD | Min. | Max. |
|---|---|---|---|---|
| **Full Models** | | | | |
| IPW for outcome Model | 0.97 | 0.66 | 0.08 | 4.70 |
| IPW for mediator Model | 0.96 | 0.39 | 0.03 | 3.33 |

## Appendix D. Code and Data Accessibility

You can find codes from https://github.com/chenglin829/SA_code
And obtain data from Sweden statistics

**List of tables and figures**

### Table 1. Overview of Measures Used for Data Analysis

| Measures | Level |
|---|---|
| **Exposure** | |
| Past social assistance recipiency (annual, 2001-2014) | Binary |
| **Outcome** | |
| Social assistance recipiency (2015 to 2017) | Binary |
| **Mediator** | |
| Neighborhood disadvantage | Continuous |
| **Time-invariant variables** | |
| Gender | |
| Female=2 | Binary |
| Male=1 | |
| Migration background | |
| Foreign-born=1 | Binary |
| Native-born=0 | |
| **Time-varying confounders** | |
| Marital status | |
| Unmarried=0 | |
| Married=1 | Categorical |
| Divorce=2 | |
| Widowed=3 | |
| Living in metropolitan county | |
| Yes=1 | Binary |
| No=0 | |
| Education | |
|   Lower than Upper secondary school | |
|   Upper secondary school | Categorical |
|   University or college | |
|   Postgraduate | |
| Residential mobility in a given year | Continuous |
| Number of children aged 7-17 | Count |
| Number of children aged 4-6 | Count |
| Number of children aged 0-3 | Count |
| Unemployment status | |
|   Yes=1 | Binary |
|   No=0 | |
| Living in non-rental housing | |
|   Yes=1 | Binary |
|   No=0 | |
| Experience of working in a skilled occupation | |
|   Yes=1 | Binary |
|   No=0 | |
| Disposable income (log) excluding social assistance | Continuous |
| Housing allowance | Continuous |
| Parental benefits | Continuous |
| Number of companies within local labor market | Continuous |
| Size of working-age population within local labor market | Continuous |

**Table 2. Overview of Time-Varying and Time-Invariant Variables**

| **Time-varying Variables** | Aged 20 (2001) | Aged 27 (2008) | Aged 33 (2014) |
|---|---|---|---|
| | Percentage | Percentage | Percentage |
| Living in non-rental housing | | | |
|   YES | 0.02 | 0.17 | 0.38 |
| NO | 0.98 | 0.83 | 0.62 |
| Registered unemployment | | | |
| YES | 0.06 | 0.09 | 0.06 |
| NO | 0.94 | 0.91 | 0.94 |
| Social assistance recipiency | | | |
| YES | 0.12 | 0.04 | 0.03 |
| NO | 0.88 | 0.96 | 0.97 |
| Experience of working in a skilled occupation | | | |
| YES | 0.03 | 0.29 | 0.35 |
| NO | 0.97 | 0.71 | 0.65 |
| Education | | | |
| Lower than Upper secondary school | 0.18 | 0.11 | 0.09 |
| Upper secondary school | 0.74 | 0.45 | 0.41 |
| University or college | 0.08 | 0.44 | 0.49 |
| Postgraduate | 0.00[a] | 0.00[b] | 0.01 |
| Marital Status | | | |
| Unmarried=0 | 0.9827 | 0.86 | 0.62 |
| Married=1 | 0.0166 | 0.13 | 0.34 |
| Divorced=2 | 0[c] | 0.01 | 0.04 |
| Widowed=3 | 0 | 0[d] | 0[e] |
| Living in metropolitan county | | | |
| YES | 0.49 | 0.55 | 0.56 |
| NO | 0.51 | 0.45 | 0.44 |

| | Mean | Sd. | Mean | Sd. | Mean | Sd. |
|---|---|---|---|---|---|---|
| Disadvantaged neighborhood score | 0.37 | 0.08 | 0.32 | 0.09 | 0.30 | 0.08 |
| Log net disposable income | 6.27 | 1.16 | 7.22 | 1.23 | 7.60 | 1.24 |
| Parental benefits (1000 kr) | 3.84 | 41.66 | 72.94 | 234.33 | 169.98 | 365.65 |
| Housing allowance (1000 kr) | 5.73 | 19.26 | 3.98 | 20.92 | 5.22 | 29.76 |
| Number of children aged 0-3 in household | 0.03 | 0.19 | 0.25 | 0.52 | 0.51 | 0.66 |
| Number of children aged 4-6 in household | 0.02 | 0.13 | 0.07 | 0.29 | 0.29 | 0.52 |
| Number of children age 7-17 in household[6] | 0.38 | 0.72 | 0.05 | 0.24 | 0.26 | 0.61 |
| Residential mobility during the year | 0.41 | 0.65 | 0.35 | 0.58 | 0.20 | 0.46 |
| Number of companies within local labor market (log) | 10.08 | 1.37 | 10.55 | 1.33 | 10.57 | 1.37 |
| Size of working-age population within local labor market (log) | 12.84 | 1.48 | 13.21 | 1.42 | 13.31 | 1.44 |
| **Time-invariant variables** | Percentage | | | | | |
| Gender | Male: 0.509 | | | | | |
| | Female: 0.491 | | | | | |
| Migration background | Foreign born: 0.101 | | | | | |
| | Native born: 0.899 | | | | | |
| Social assistance recipiency at age 34-36 | Yes: 0.03 | | | | | |
| | No: 0.97 | | | | | |

*Note*:

[a] Not actually 0 but 0.0003

[b] Not actually 0 but 0.0005

[c] Not actually 0 but 0.0007

[d] Not actually 0 but 0.00008

[e] Not actually 0 but 0.0005

Table 3. Effect Size of MSM Estimates

| Models | Coef. | SE |
|---|---|---|
| **Full Models** | | |
| Outcome Model $\theta_1$ | 0.342*** | 0.006 |
| Outcome Model $\theta_2$ | 0.229*** | 0.031 |
| Mediator Model $\beta_1$ | 0.159*** | 0.002 |

*Note*: *p < 0.05. **p < 0.01. ***p < 0.001(two-tailed tests).

Table 4. Robustness Check Models

| Models | IDE $T\theta_1$ | IIE $T\beta_1\theta_2$ | IDE+IIE | % Mediated |
|---|---|---|---|---|
| Model 0 | 4.45 | 0.47 | 4.92 | 9.5% |
| Model 1 (k=500) | 4.59 | 0.26 | 4.85 | 5.4% |
| Model 2 (1983 cohort) | 4.38 | 0.35 | 4.73 | 7.4% |
| Model 3 (Outcome=1 if at least 10% of disposable income is social assistance) | 5.07 | 0.43 | 5.60 | 7.8% |

*Note*: *The coefficients are expressed on a risk ratio scale.*

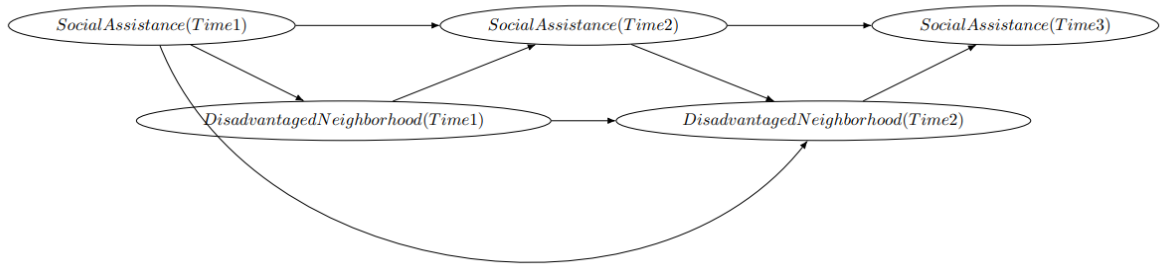

*Note*: Social Assistance indicates social assistance recipiency. Disadvantaged Neighborhood indicates exposure to a disadvantaged neighborhood.
**Figure 1. Three-Period Modification of Kuh's (2014) *Chain of Risk Additive Model***

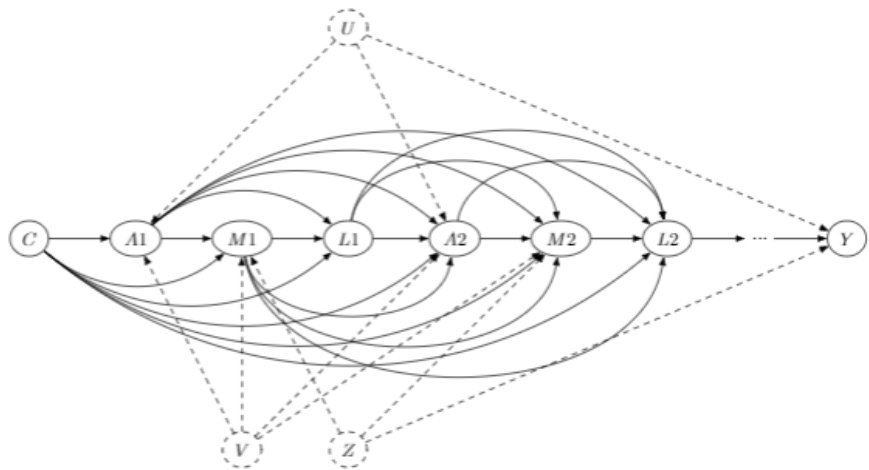

*Note: C:* Baseline covariates; A: Prior social assistance recipiency (Treatment); M: Disadvantage level of neighborhood (Mediator); L: Time-varying confounders; Y: Social assistance recipiency (end-of-study outcome). The figure is adapted from VanderWeele and Tchetgen Tchetgen (2017). U, V, and Z are unobservable confounders for A-Y, A-M, and M-Y respectively.

**Figure 2. Directed Acyclic Graph (DAG)**

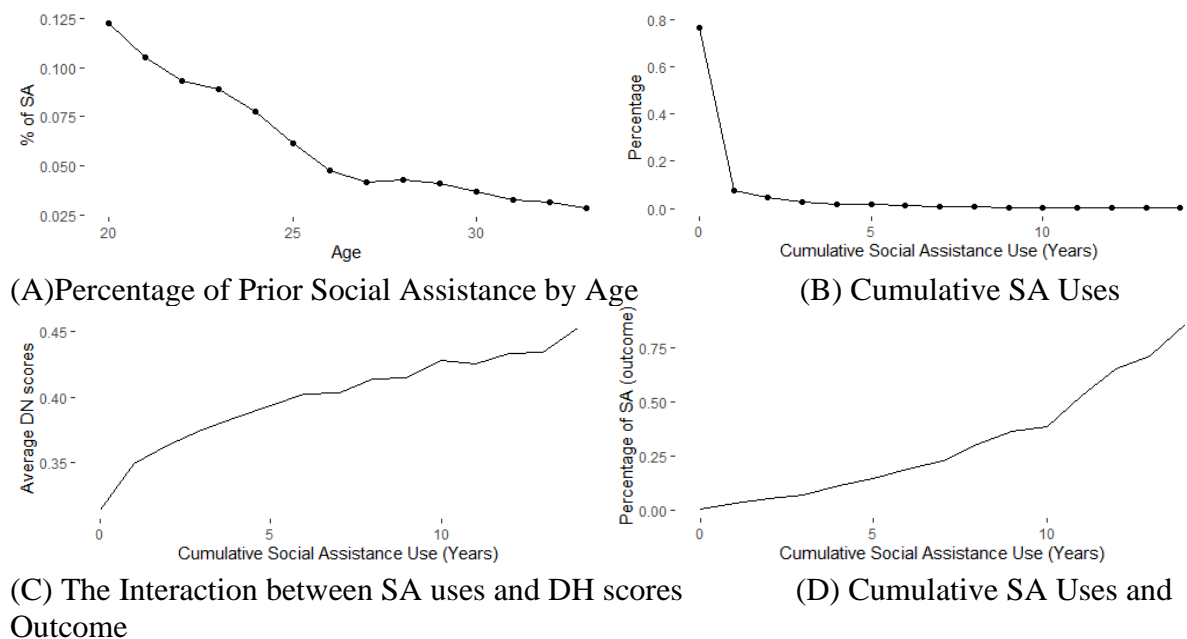

(A) Percentage of Prior Social Assistance by Age

(B) Cumulative SA Uses

(C) The Interaction between SA uses and DH scores Outcome

(D) Cumulative SA Uses and

**Figure 3. Patterns of Social Assistance Recipiency**

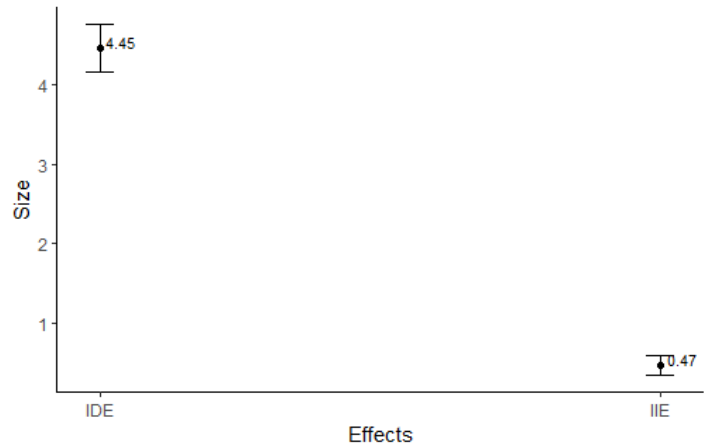

*Note*:
The coefficients are expressed on a risk ratio scale.
**Figure 4. Interventional Effects**

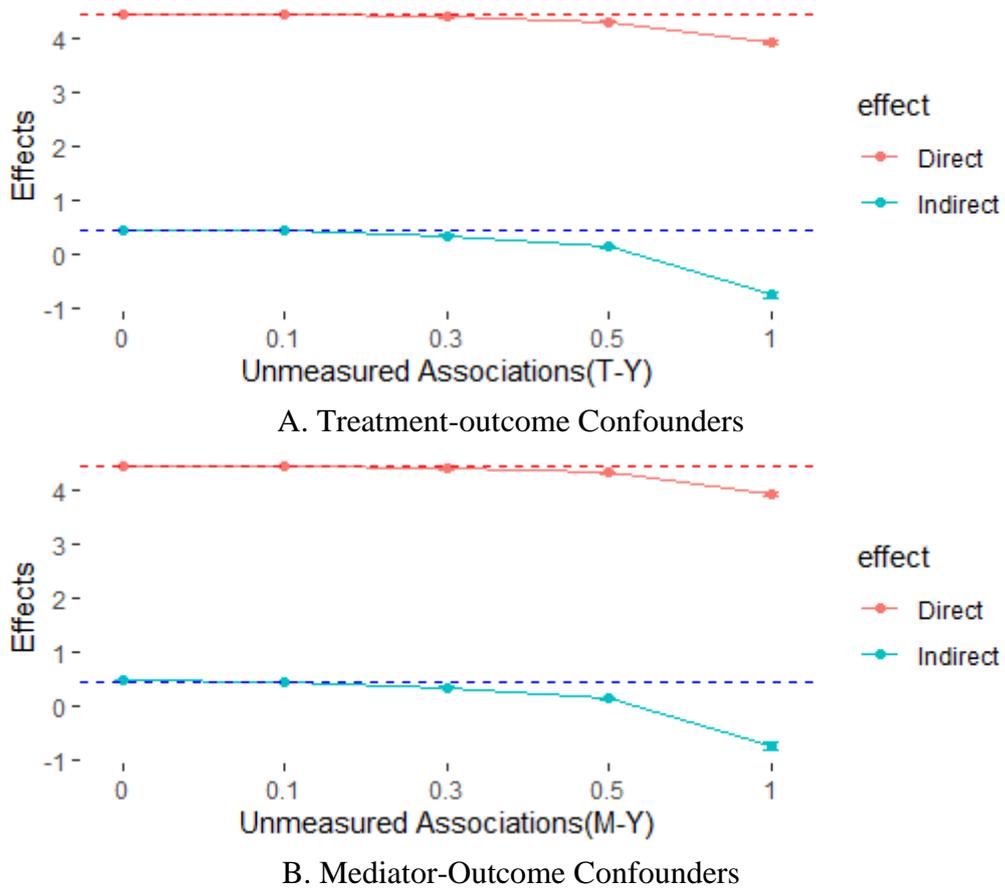

A. Treatment-outcome Confounders

B. Mediator-Outcome Confounders

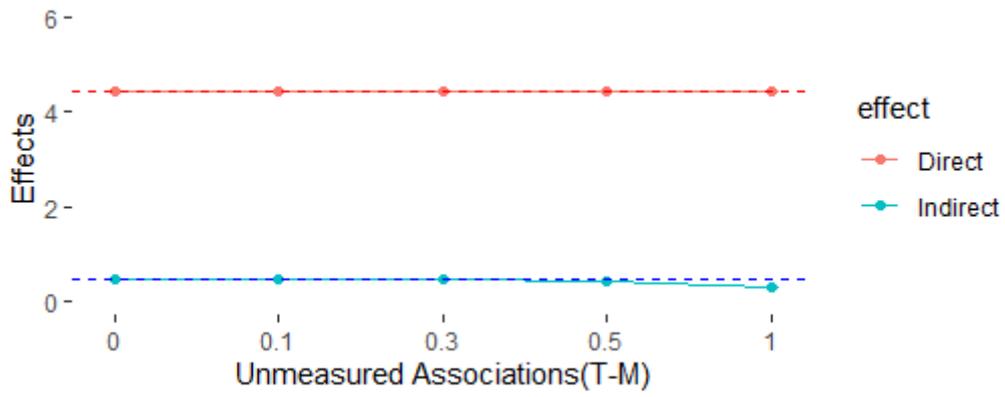

C. Treatment-mediator Confounders

*Note*:
Blued dashed line: Estimated interventional indirect effect
Red dashed line: Estimated interventional direct effect
X-axis: The potential size of associations between unmeasured confounders between treatment-and-outcome, mediator-and-outcome and treatment-and-mediator respectively.

**Figure 5. Sensitivity Analyses**